\documentclass[11pt,letterpaper]{JHEP3}
\usepackage{amsmath,epsfig,array}
\usepackage{amssymb}
\usepackage{amsfonts,amstext,latexsym,xspace}
\usepackage{amsfonts}
\newcommand{\N}{{\mathcal N}}

\newcommand{\eq}{\begin{equation}}
\newcommand{\en}{\end{equation}}
\newcommand{\eqn}{\begin{eqnarray}}
\newcommand{\enn}{\end{eqnarray}}
\newcommand{\nn}{\nonumber }
\newcommand{\beq}{\begin{equation}}
\newcommand{\eeq}{\end{equation}}
\newcommand{\M}{\ensuremath{\mathcal{M}}}

\newcommand{\gd}[1][{}]{\delta_{#1}{}}

\newcommand{\2}{{\textstyle {\frac{1}{2}} }}

\renewcommand{\i}{\mathrm{i}}

\newcommand{\cX}{\mathcal{X}}
\newcommand{\cY}{\mathcal{Y}}

\newcommand{\cN}{\mathcal{N}}

\newcommand{\FTP}[4][{}]{\left( #2,#3,#4 \right)^{#1}}

\newcommand{\pa}{\partial}
\newcommand{\lpr}{{\cal L}^{(+4)}}

\newcommand{\fracds}[2]{\displaystyle \frac{#1}{#2} }
\def\a{\alpha} \def\b{\beta} \def\e{\epsilon}

\def\G{\Gamma} 

\title{Lectures on Spectrum Generating Symmetries and U-duality in Supergravity , Extremal Black Holes, Quantum Attractors and Harmonic Superspace}
\author{
Murat G\"{u}naydin\footnote{murat@phys.psu.edu}
\\
\emph{Institute for Gravitation and the Cosmos \\ Physics Department \\
Penn  State University\\
University Park, PA 16802, USA} \\
}

\abstract{We  
review the underlying algebraic structures of supergravity theories with symmetric scalar manifolds in five and four dimensions,    orbits of their extremal black hole solutions  and  the spectrum generating extensions of their U-duality  groups.  For $5D$, $N=2$ Maxwell-Einstein supergravity theories (MESGT) defined by Euclidean Jordan algebras ,$J$, the spectrum generating symmetry groups  are the conformal groups $Conf(J)$ of $J$ which are  isomorphic to their U-duality groups in four dimensions.  Similarly, the  spectrum generating symmetry groups of  $4D$, $N=2$  MESGTs are the quasiconformal groups $QConf(J)$ associated with $J$ that are isomorphic to their  U-duality groups in three dimensions.   We  then review the work on  spectrum generating symmetries  of spherically symmetric stationary $4D$ BPS  black holes, based on the equivalence of their attractor  equations  and the equations for geodesic motion of a fiducial particle on the target spaces of   corresponding $3D$ supergravity theories obtained by timelike reduction.  We  also  discuss the connection between  harmonic superspace formulation of $4D$ , $N=2$ sigma models coupled to supergravity and the minimal unitary representations of their isometry groups obtained by quantizing their quasiconformal realizations.  We discuss the relevance of this connection  to spectrum generating symmetries and  conclude with a brief summary  of  more recent  results.}

\preprint{}

\begin{document}
\renewcommand{\theequation}{\arabic{section}.\arabic{equation}}
\section{Introduction}
This review on spectrum generating symmetries in supergravity , extremal black holes, U-duality orbits, quantum attractor flows  and harmonic superspace  is based on four lectures given at the School on Attractor Mechanism (SAM 2007) in Frascati, Italy. The lectures were titled \\
{\it  1. Very special real geometry , Jordan algebras and attractors \\
2. Very special complex geometry , Freudenthal triple systems and attractors \\
3. Conformal and quasiconformal extensions of U-duality groups as spectrum generating symmetry groups \\
4. Harmonic superspace, quasiconformal groups and their minimal unitary representations. } \\
I will follow closely the material covered in the lectures and discuss briefly the results obtained since SAM 2007 at the end. 

More specifically, section 2 is a review of the U-duality symmetries of maximal supergravity in various dimensions.  In section 3 and 4  we review the $N=2$ Maxwell-Einstein supergravity theories (MESGT) in five dimensions and the connection between Jordan algebras of degree three and the  MESGTs with symmetric scalar manifolds. This is followed by a review of the symmetry groups of Jordan algebras using the language of space-time symmetry groups and a complete list of simple finite dimensional Jordan algebras and their automorphism ( rotation), reduced structure ( Lorentz) and M\"obius ( conformal) groups.
In section 6 we review the U-duality orbits of extremal black holes of $5D$ supergravity theories with symmetric target manifolds and discuss how these results lead to the proposal that
$4D$ U-duality groups act as spectrum generating symmetry groups of  the corresponding $5D$ theories. Section 7 is an overview of the $4D$ MESGTs with symmetric target spaces and their connection with Freudenthal triple systems. 
This is followed by a classification of the U-duality orbits of extremal black holes of $N=2$ MESGTs with symmetric target manifolds and of $N=8$ supergravity 
and the proposal that the three dimensional U-duality groups  act as spectrum generating quasiconformal groups of the corresponding $4D$ theories. 
In section 9 we summarize the novel quasiconformal realizations of non-compact groups and their relation to Freudenthal triple systems. A precise and concrete  implementation of the proposal that $3D$ U-duality groups act as spectrum generating symmetry groups of the corresponding $4D$ theories within the framework of spherically symmetric stationary BPS black holes is discussed in section 10.  We then review the connection between the harmonic superspace formulation of $4D$ , $N=2$ sigma models coupled to supergravity and the minimal unitary representations of their isometry groups. For sigma models with symmetric target spaces we show that there is a remarkable map between the Killing potentials that generate their isometry groups in harmonic superspace and the minimal unitary representations of these groups obtained by quantizing their quasiconformal realizations.  Implications of this result are also discussed. Section12 is devoted to the M/superstring theoretic origins of $N=2$ MESGTs with symmetric target spaces, in particular , the magical supergravity theories. We conclude with a brief discussion of the  related developments that took place since SAM 2007 and some open problems. 
\section{U-duality Symmetries of Maximal Supergravity  in various Dimensions}
The maximal possible dimension for Poincare supergravity is eleven dimensions  \cite{Nahm:1977tg}. Eleven dimensional supergravity  involves a Majorana gravitino field, the elf-bein $E_M^A$  and an anti-symmetric tensor of rank three $A_{MNP}$ and was constructed in \cite{Cremmer:1978km}.
Lagrangian of its bosonic sector has a very simple form
\begin{eqnarray}
{\cal L}_{11}&=& {1\over \kappa_{11}^2} \bigg[
-\frac{1}{2} E\, R
-\frac{1}{48}E
(F_{MNPQ})^2 \nonumber\\
&& - \frac{\sqrt{2}}{3456}  \, \varepsilon^{MNPQRSTUVWX}
\,F_{MNPQ} \,F_{RSTU} \,A_{VWX} + \cdots\bigg]
\end{eqnarray}
where the ellipses denote terms that involve fermions, $R$ is the scalar curvature, $E$ is the determinant of 11 dimensional elf-bein $E_M^A$ and $F_{MNPQ}$ is the field strength of the antisymmetric three-form field $A_{MNP}$ with $M,N,P,...= 0,1,....,10$.

Under dimensional reduction to $d$ dimensions, eleven dimensional supergravity results in the maximal supergravity with a global symmetry group $E_{((11-d)(11-d))}$. We shall refer to these global symmetry groups as the U-duality groups even though the term was originally used for discrete subgroups of these continuous groups which are believed to be non-perturbative  symmetries of $M$-theory toroidally compactified to $d$ dimensions \cite{Hull:1994ys}.

In six dimensions maximal supergravity  has the global symmetry group $E_{5(5)}= SO(5,5)$ and the scalar fields of this theory parametrize the symmetric space
\eq
\mathcal{M}_6 = \frac{SO(5,5)}{SO(5)\times SO(5)}
\en
In five dimensions the global symmetry group of maximal supergravity is $E_{6(6)}$ under which all  the vector fields transform in the irreducible 27 dimensional representation of $E_{6(6)}$.
The scalar fields of the theory parametrize the symmetric space
\eq
\mathcal{M}_5 = \frac{E_{6(6)}}{USp(8)}
\en
U-duality group $E_{6(6)}$ is a symmetry of the Lagrangian of ungauged maximal supergravity in five dimensions.

In four dimensions the maximal ungauged supergravity has the U-duality group $E_{7(7)}$ as an {\it on-shell symmetry} under which  field strengths of the 28 vector fields of the theory and their magnetic duals transform in the 56 dimensional representation , which is the smallest non-trivial representation of $E_{7(7)}$. The 70 scalar fields of the theory parametrize the symmetric space
\eq
\mathcal{M}_4 =\frac{E_{7(7)}}{SU(8)}
\en
{\it Symmetry of the Lagrangian } of ungauged $N=8$ supergravity theory in four dimensions depends on the real symplectic section chosen. Dimensional reduction from 5 dimensions leads to a symplectic section with $E_{6(6)} \times SO(1,1)$ symmetry of the Lagrangian such that under $E_{6(6)}$ electric field strengths transform in the reducible representation $(27+1)$. There exist also symplectic sections leading to Lagrangians with $SL(8,\mathbb{R})$  and $SU^*(8)$ symmetry groups. In the latter two cases the electric field strengths transform irreducibly in the 28 dimensional real representation of $SL(8,\mathbb{R})$ or $SU^*(8)$, respectively.

In three dimensions all the dynamical bosonic degrees of freedom of  maximal supergravity can be dualized to scalar fields parametrizing the symmetric space \cite{Marcus:1983hb}
\eq
\mathcal{M}_3 = \frac{E_{8(8)}}{SO(16)}
\en

\section{$5D$, $N=2$ Maxwell-Einstein Supergravity Theories}
\setcounter{equation}{0}
Certain matter coupled supergravity theories do admit global symmetry groups which we shall also refer to as U-duality groups. In this section we shall study the U-duality groups that arise in five dimensional $N=2$ Maxwell-Einstein supergravity theories (MESGT). Five dimensional MESGTs that describe the coupling of an arbitrary number
of $N=2$ (Abelian) vector multiplets to $N=2$ supergravity were constructed  long ago
in \cite{Gunaydin:1983rk,Gunaydin:1983bi,Gunaydin:1984ak,Gunaydin:1986fg}.
The fields of the graviton supermultiplet are the f\"unfbein $e_\mu^m$ , two gravitini $\psi_\mu^i$ $(i=1,2)$ , and a vector field $A_\mu$ ( the "bare" graviphoton). A vector multiplet consists of a vector field $A_\mu$, two "gaugini" $\lambda^i$ and one real scalar $\phi$.
The bosonic part of five dimensional  $N=2$ MESGT Lagrangian describing the coupling of $(n_V-1)$ vector multiplets has a very simple form \footnote{%
We use the conventions of \cite{Gunaydin:1983bi}.}
\begin{eqnarray}\label{Lagrange}
e^{-1}\mathcal{L}_\textrm{bosonic} &=& -\frac{1}{2} R
-\frac{1}{4}{\stackrel{\circ}{a}}_{IJ} F_{\mu\nu}^{I} F^{J\mu\nu}-
 \frac{1}{2} g_{xy}(\partial_{\mu}\varphi^{x})
(\partial^{\mu} \varphi^{y})+\nonumber \\
 && + \frac{e^{-1}}{6\sqrt{6}} C_{IJK} \varepsilon^{\mu\nu\rho\sigma\lambda}
 F_{\mu\nu}^{I}F_{\rho\sigma}^{J}A_{\lambda}^{K},
\end{eqnarray}
where
\begin{eqnarray*}
I&=& 1,\ldots, n_V\\
x&=& 1,\ldots, (n_V-1)\\
\mu, \nu,... &=& 0,1,2,3,4.
\end{eqnarray*}
Note that we combined the "bare" graviphoton" with the other vector fields and labelled them with a single index $I$ which runs from 1 to $n_V$. $e$ and $R$ denote the f\"{u}nfbein determinant and
scalar curvature of spacetime, respectively. $F_{\mu\nu}^{I}$ are field strengths
of the vector fields $A_{\mu}^{I}$. The metric, $g_{xy}$, of the
scalar manifold $\M_5$ and the "metric" ${\stackrel{\circ}{a}}_{IJ}$ of the kinetic energy term of vector fields
both depend on the scalar fields $\varphi^{x}$. On the other hand, the
completely symmetric tensor $C_{IJK}$ is constant as required by local
Abelian gauge symmetries of vector fields.

Remarkably,  the entire $5D$,
$N=2$ MESGT is uniquely determined by the constant tensor $C_{IJK}$
\cite{Gunaydin:1983bi}. In particular, geometry of the scalar manifold
$\mathcal{M}_5$ is determined by $C_{IJK}$ as follows. One
defines a cubic polynomial, $\mathcal{V}(h)$, in $n_V$ real variables
$h^{I}$ $(I=1,\ldots,n_V)$ using the C-tensor,
\begin{equation}
 \mathcal{V}(h):=C_{IJK} h^{I} h^{J} h^{K}\ .
\end{equation}
and a metric, $a_{IJ}$, of a $n_V$ dimensional ambient space
$\mathcal{C}_{n_V}$ coordinatized by $h^{I}$:
\begin{equation}\label{aij}
  a_{IJ}(h):=-\frac{1}{3}\frac{\partial}{\partial h^{I}}
  \frac{\partial}{\partial h^{J}} \ln \mathcal{V}(h)\ .
\end{equation}
It can then be proven that the $(n_V-1)$-dimensional manifold, $\mathcal{M}_5$, of scalar fields
$\varphi^{x}$ can  be represented as an hypersurface defined by
the condition \cite{Gunaydin:1983bi}
\begin{equation}\label{hyper1}
{\cal V} (h)=C_{IJK}h^{I}h^{J}h^{K}=1 \ ,
\end{equation}
in this ambient space $\mathcal{C}_{n_V}$\footnote{ The ambient space
$\mathcal{C}_{n_V}$ is the 5 dimensional counterpart of the hyper
K\"ahler cone of the twistor space of the corresponding three
dimensional quaternionic geometry of the scalar manifold
$\mathcal{M}_3$.}.   The metric $g_{xy}$ is simply the pull-back of
(\ref{aij}) to $\mathcal{M}_5$:
\begin{equation}
g_{xy}(\varphi)= \left(h^{I}_{x}  h^{J}_{y} a_{IJ}\right)|_{\mathcal{V}=1}
\end{equation}
where \[ h^{I}_{x} = \sqrt{\frac{3}{2}}
\frac{\partial h^I }{\partial\phi^{x}}|_{\mathcal{V}=1} \] and the ``metric''
${\stackrel{\circ}{a}}_{IJ}(\varphi)$ of kinetic energy term of the
vector fields is given by the componentwise restriction of the metric
$a_{IJ}$ of the ambient space $\mathcal{C}_{n_V}$ to $\mathcal{M}_5$:
\begin{equation*}
{\stackrel{\circ}{a}}_{IJ}(\varphi)=a_{IJ}|_{{\cal V}=1} \ .
\end{equation*}
Riemann curvature tensor of the scalar manifold takes on a very simple form
\begin{equation}
  K_{xyzu}= \frac{4}{3} \left( g_{x[u} g_{z]y} + {T_{x[u}}^{w} T_{z]yw} \right)
\end{equation}
where $T_{xyz}$ is a symmetric tensor that is the pull-back of the C-tensor to the hypersurface
\begin{equation}
   T_{xyz}= h^{I}_{x }h^{J}_{y} h^{K}_{z} C_{IJK}
\end{equation}
Since the  Riemann curvature tensor $K_{xyzu}$ depends only on the metric $g_{xy}$ and the tensor $T_{xyz}$ it follows
that the covariant constancy of $T_{xyz}$ implies the covariant
constancy of $K_{xyzu}$:
\begin{equation*}
T_{xyz ; w} = 0
\longrightarrow
 K_{xyzu ; w} =0
\end{equation*}
Therefore scalar manifolds $\mathcal{M}_5$ with covariantly constant
$T$ tensors are locally symmetric spaces.
If the scalar manifold $\mathcal{M}_5$ is  homogeneous then the covariant constancy of
$T_{xyz}$ is equivalent to the ``adjoint
identity'' for the C-tensor~\cite{Gunaydin:1983bi}:
\begin{equation}
   C^{IJK} C_{J(MN} C_{PQ)K} = \delta^{I}_{(M} C_{NPQ)}
\end{equation}
where the indices are raised by the inverse
${\stackrel{\circ}{a}}{}^{IJ}$ of ${\stackrel{\circ}{a}}{}_{IJ}$.
Furthermore, cubic forms defined by $C_{IJK}$ of $N=2$ MESGT's that
satisfy the adjoint identity and lead to positive definite metrics $g_{xy}$ and  ${\stackrel{\circ}{a}}_{IJ}(\varphi)$ are in one-to-one correspondence with
norm forms of Euclidean (formally real) Jordan algebras $J$ of degree 3
\cite{Gunaydin:1983bi}. The scalar manifolds of the corresponding MESGT's are of the form
\eq
\mathcal{M}_5 = \frac{Str_0(J)}{Aut(J)}
\en
where $Str_0(J)$ is the invariance group of the norm $\mathbf{N}$ of $J$ and $Aut(J)$ is its automorphism group.
 These theories exhaust the list of $5D$
MESGTs with symmetric target spaces $G/H$ such that $G$ is a symmetry
of the Lagrangian \cite{deWit:1992wf}.  Remarkably, the list of
cubic forms that satisfy the adjoint identity coincides also with the list
of Legendre invariant cubic forms that were classified more recently by mathematicians
\cite{MR2094111}.
Before we discuss the geometries of $N=2$ MESGT's defined by Jordan algebras we shall take a detour and review some of the basic facts regarding Jordan algebras of degree three in the next subsection.
For details and further references on Jordan algebras we refer to the monograph  \cite{MR2014924}

\section{ MESGT's with Symmetric Target Spaces and Euclidean  Jordan Algebras of Degree Three}

A Jordan algebra $J$ over a field $\mathbb{F}$ is a commutative and non-associative  algebra
with a product $\circ$ that satisfies
\begin{equation}\label{commute} X\circ Y = Y
\circ X \in J, \quad \forall\,\, X,Y \in J \ ,
\end{equation}
and
\begin{equation}\label{Jidentity}
X\circ (Y \circ X^2)= (X\circ Y) \circ X^2 \ ,
\end{equation}
where $X^2\equiv (X\circ X)$.  Given a Jordan algebra $J$, one can define a norm form
\[ \mathbf{N}:J\rightarrow \mathbb{R}\]  that satisfies the composition
property \cite{MR0251099}
\begin{equation}\label{Norm}
\mathbf{N}[2X\circ(Y\circ X)-(X\circ X)\circ Y]=\mathbf{N}^{2}(X) \mathbf{N}(Y).
\end{equation}
A Jordan algebra is said to be of degree, $p$, if its norm form satisfies
$\mathbf{N}(\lambda X)=\lambda^p \mathbf{N}(X)$, where $\lambda\in \mathbb{R}$.  A
\emph{Euclidean} Jordan algebra is a Jordan algebra for which the
condition $X\circ X + Y\circ Y=0$ implies that $X=Y=0$ for all $X,Y\in
J$. Euclidean Jordan algebras are sometimes called compact Jordan algebras since their
automorphism groups are compact.

As explained above, given a Euclidean Jordan
algebra of degree three one can identify its norm form $\mathbf{N}$ with the
cubic polynomial $\mathcal{V}$ defined by the C-tensor of a 5D, $N=2$
MESGT with a symmetric scalar manifold \cite{Gunaydin:1983bi}.
 Euclidean Jordan algebras of degree three fall into an infinite family of non-simple Jordan algebras which are direct sums of the form
\begin{equation}
  J=\mathbb{R}\oplus \Gamma_{(1,n-1)}
\end{equation}
where $\Gamma_{(1,n-1)}$ is an $n$
dimensional Jordan algebra of degree two associated with a quadratic
norm form in $n$ dimensions that has a ``Minkowskian signature''
$(+,-,\ldots,-)$ and $\mathbb{R}$ is the one dimensional Jordan algebra.
 This infinite family of reducible Jordan algebras of degree 3 exists for any $n$ and  is referred to as  the generic
Jordan family.
 The scalar manifolds of corresponding 5D, $N=2$ MESGT's are
\begin{equation}
 \mathcal{M}_5(\mathbb{R}\oplus \Gamma_{(1,n-1)}) =\frac{SO(n -1,1)}{SO(n -1)}\times  SO(1,1)
\end{equation}

  An irreducible realization of $\Gamma_{(1,n-1)}$ is provided by
$(n-1)$ ($2^{[n/2]} \times 2^{[n/2]} $) Dirac gamma matrices $\gamma^i$ $(i,j,\ldots=1,\ldots,(n-1))$
of an $(n-1)$ dimensional Euclidean space together with the identity
matrix $ \gamma^0 = \mathbf{1}$ and the Jordan product $\circ$ being
defined as one half the anticommutator:

\begin{eqnarray}
\gamma^i \circ \gamma^j &=& \frac{1}{2}
\{\gamma^i,\gamma^j\}= \delta^{ij} \gamma^0
\nonumber \\
\gamma^0 \circ \gamma^0 &=& \frac{1}{2}
\{\gamma^0,\gamma^0\}= \gamma^0 \nonumber\\
\gamma^i \circ \gamma^0 &=& \frac{1}{2}
\{\gamma^i,\gamma^0\}= \gamma^i \ .
\end{eqnarray}
  The quadratic norm of a general element $\mathbb{X} = X_0 \gamma^0 + X_i \gamma^i $ of
$\Gamma_{(1,n-1)}$
is defined as
\begin{equation*}
  \mathbf{Q}(\mathbb{X}) = \frac{1}{2^{[n/2]}} \mathop{Tr}
  \mathbb{X} \bar{\mathbb{X}} = X_0X_0 - X_iX_i \ ,
\end{equation*}
where
\begin{equation*}  \bar{\mathbb{X}} \equiv  X_0 \gamma^0 - X_i \gamma^i  \ .
\end{equation*}
The norm of a general element $y \oplus \mathbb{X} $ of the non-simple
Jordan algebra $J=\mathbb{R}\oplus \Gamma_{(1,n-1)}$ is then
simply
\begin{equation}
  \mathbf{N}(y \oplus \mathbb{X}) =  y \mathbf{Q}(\mathbb{X})
\end{equation}
where $y\in \mathbb{R}$.

In addition to this generic reducible infinite family, there exist four simple Euclidean Jordan algebras of degree three. They are realized  by Hermitian $(3\times 3)$-matrices over the four
division algebras $\mathbb{A} = \mathbb{R}, \mathbb{C},
\mathbb{H}, \mathbb{O}$ ( reals $\mathbb{R}$ , complex numbers $\mathbb{C}$ , quaternions 
$\mathbb{H}$  and octonions  $\mathbb{O}$.)
 \begin{equation}
 J =
\left(
  \begin{array}{ccc}
    \alpha & Z & \bar{Y} \\
    \bar{Z} & \beta & X \\
    Y & \bar{X} & \gamma \\
  \end{array}
\right)
\end{equation}
where $\alpha , \beta, \gamma \in \mathbb{R}$ and $X,Y,Z \in
\mathbb{A}$ with the Jordan product being one half the anticommutator.  They
are denoted as $J_3^{\mathbb{R}}$, $J_3^{\mathbb{C}}$,
$J_3^{\mathbb{H}}$, $J_3^{\mathbb{O}}$, respectively.  The
corresponding $N=2$ MESGTs are called ``magical supergravity
theories''\cite{Gunaydin:1983rk}.  The scalar manifolds of the magical supergravity theories in five dimensions are the irreducible symmetric spaces
\begin{eqnarray}\label{magicals}
J_{3}^{\mathbb{R}}:\quad \mathcal{M}&=& SL(3,\mathbb{R})/
SO(3)\qquad
\nonumber\\
J_{3}^{\mathbb{C}}:\quad \mathcal{M}&=& SL(3,\mathbb{C})/
SU(3)\qquad
\nonumber\\
J_{3}^{\mathbb{H}}:\quad \mathcal{M}&=& SU^{*}(6)/
USp(6)\qquad
 \nonumber \\
J_{3}^{\mathbb{O}}:\quad \mathcal{M}&=& E_{6(-26)}/
F_{4}\qquad \qquad
\end{eqnarray}

The cubic norm form, $\mathbf{N}$, of the simple Jordan algebras is given by the
determinant of the corresponding Hermitian $(3\times 3)$-matrices.
\begin{equation}
\mathbf{N}(J)= \alpha \beta \gamma - \alpha X \bar{X} - \beta Y \bar{Y} - \gamma Z \bar{Z} + 2 Re (X Y Z )
\end{equation}
$Re(XYZ)$ denotes the real part of $XYZ$
\eq
Re(XYZ) = Re(X(YZ))=Re((XY)Z)= \frac{1}{2} \left(XYZ +\overline{XYZ} \right)
\en
where bar denotes conjugation in the underlying division algebra.

A real quaternion $X \in \mathbb{H}$ can be expanded as
\begin{eqnarray}
X&=&X_0 + X_1 j_1 + X_2 j_2 + X_3 j_3 \nonumber \\
\bar{X}& = &X_0 - X_1 j_1 - X_2 j_2 - X_3 j_3 \\
X \bar{X}& =& X_0^2 + X_1^2 + X_2^2 +X_3^2 \nonumber
\end{eqnarray}
where the imaginary units $j_i \hspace{.25cm} (i=1,2,3)$ satisfy
\begin{equation}
j_i j_j = - \delta_{ij} + \epsilon_{ijk} j_k
\end{equation}
A real  octonion $X\in \mathbb{O}$ has an expansion
\begin{eqnarray}
X &=& X_0 + X_1 j_1 + X_2 j_2 + X_3 j_3  + X_4 j_4 + X_5 j_5  +X_6 j_6 + X_7 j_7 \nonumber \\
\bar{X} &=& X_0 - X_1 j_1 - X_2 j_2 - X_3 j_3  - X_4 j_4 - X_5 j_5  - X_6 j_6 - X_7 j_7 \\
X \bar{X} &= &X_0^2 + \sum_{A=1}^7 (X_A)^2 \nonumber
\end{eqnarray}
where the seven imaginary units  $j_A\hspace{.25cm} ( A=1,2,...,7 )$ satisfy
\begin{equation}
j_A j_B = - \delta_{AB} + \eta_{ABC} j_C
\end{equation}
The $G_2$ invariant tensor $\eta_{ABC}$ is completely antisymmetric and in the conventions of \cite{Gunaydin:1973rs} its nonvanishing components  take  on the values
\begin{equation}
\eta_{ABC} = 1  \Leftrightarrow  (ABC)= (123), (471), (572), (673), (624), (435), (516)
\end{equation}

The bosonic content and scalar manifold of $N=6$ supergravity is the same as that of the $N=2$
MESGT defined by the simple Euclidean Jordan algebra
$J_3^{\mathbb{H}}$\cite{Gunaydin:1983rk}, namely
\begin{equation} \mathcal{M}_5 = \frac{SU^*(6)}{USp(6)} \end{equation}
Therefore its invariant C-tensor is simply the one given by the cubic
norm of $J_3^{\mathbb{H}}$.

The C-tensor $C_{IJK}$ of  $N=8$  supergravity in five dimensions can be identified with the
symmetric tensor given by the cubic norm of the split exceptional Jordan algebra
$J_3^{\mathbb{O}_s}$ \cite{Ferrara:1997uz,Gunaydin:2000xr} defined over split octonions $\mathbb{O}_s$. The automorphism group of the split exceptional Jordan
algebra defined by $ 3\times 3 $ Hermitian matrices over the split octonions $\mathbb{O}_s$
\begin{equation}
J^s =\left(
       \begin{array}{ccc}
         \alpha & Z^s& \bar{Y}^s\\
         \bar{Z}^s & \beta & X^s \\
         Y^s & \bar{X}^s & \gamma \\
       \end{array}
     \right)
\end{equation}
,where $X^s,Y^s,Z^S$ are split octonions,  is the noncompact group $F_{4(4)}$ with the maximal compact subgroup $USp(6)\times USp(2)$. Its reduced structure  group is $E_{6(6)}$ under which the C-tensor is invariant. $E_{6(6)}$ is  the U-duality  group of maximal supergravity  in five dimensions whose  scalar manifold is
\begin{equation*} E_{6(6)} /USp(8) \end{equation*}

\section{ Rotation ( Automorphism), Lorentz (Reduced Structure) and Conformal (M\"obius) Groups
of Jordan Algebras}

Above we reviewed briefly the  connections between  Jordan algebras of degree three and supergravity
 theories.  Jordan algebras were used in the very early days of spacetime supersymmetry
to define generalized spacetimes that naturally extend the description of four dimensional Minkowski spacetime  and its symmetry
groups in terms of $2\times 2$ complex Hermitian matrices. This was mainly motivated by  attempts to find the super analogs of the
exceptional Lie algebras \cite{Gunaydin:1975mp} before a complete classification of finite dimensional simple Lie superalgebras was given by Kac \cite{Kac:1977em}.

As is well-known the  twistor formalism in four-dimensional
space-time $(d=4)$ leads naturally to the representation of spacetime coordinates $x_{\mu}$ in terms of $2\times 2$ Hermitian matrices over the field of
complex numbers ${\mathbb C}$:
\begin{equation}
    x=x_{\mu}\sigma^{\mu}
\end{equation}
Hermitian matrices over the field of complex numbers close
under the symmetric anti-commutator product and form a simple Jordan algebra denoted as $J_2^{\mathbb C}$. Therefore one can regard the four dimensional Minkowski 
coordinate vectors as elements of the  Jordan algebra
$J_2^{\mathbb C}$ \cite{Gunaydin:1975mp,Gunaydin:1979df}.  Then the rotation, Lorentz and
conformal groups in four dimensions correspond simply to  the automorphism,
reduced structure and M\"{o}bius (linear fractional) groups of the
Jordan algebra $J_2^{\mathbb
C}$ \cite{Gunaydin:1975mp,Gunaydin:1979df}.  The reduced structure group $Str_0(J)$ of a
Jordan algebra $J$ is simply the invariance group of its norm form
$\mathbf{N}(J)$, while the structure group $Str(J)= Str_0(J)\times SO(1,1)$ is defined as  the invariance group of $\mathbf{N}(J)$ modulo an overall nonzero global scale factor. This correspondence was then used to  define generalized space-times coordinatized by  elements of general Jordan  algebras  whose
rotation $Rot(J)$, Lorentz $Lor(J)$ and conformal $Conf(J)$ groups are  identified with the
automorphism $Aut(J)$, reduced structure $Str_0(J)$ and M\"obius M{\"o}b$(J)$ groups of $J$
\cite{Gunaydin:1975mp,Gunaydin:1979df,Gunaydin:1989dq,Gunaydin:1992zh}.
 Denoting as $J_{n}^{\mathbb A}$ the
Jordan algebra of $n\times n$ Hermitian matrices over the {\it
division} algebra ${\mathbb A}$ and the Jordan algebra of Dirac gamma
matrices in $d$ (Euclidean) dimensions as $\Gamma_{(1,d)}$ we list the
 symmetry groups of generalized space-times defined by simple
Euclidean (formally real) Jordan algebras in Table \ref{euclideanjordanlist}

\begin{table}
\begin{center}
\begin{tabular}{|c|c|c|c|}
 \hline
$J $& $Rot(J)$ & $Lor(J)$ & $Conf(J)$ \\
\hline
~&~&~&~\\
$J_{n}^{\mathbb R}$ & $SO(n)$ & $SL(n,{\mathbb R})$ & $Sp(2n,{\mathbb R})$\\
~&~&~&~\\
$J_{n}^{\mathbb C}$ & $SU(n) $ & $ SL(n,{\mathbb C})$ & $SU(n,n)$ \\
~&~&~&~\\
$J_{n}^{\mathbb H} $& $USp(2n)$ & $SU^{*}(2n)$ &$ SO^{*}(4n)$ \\
~&~&~&~\\
$J_{3}^{\mathbb O}$ &$ F_{4}$ & $E_{6(-26)}$ & $E_{7(-25)}$ \\
~&~&~&~ \\
$\Gamma_{(1,d)}$ & $SO(d)$ & $SO(d,1)$ & $SO(d,2)$ \\
~&~&~&~ \\
\hline
\end{tabular}

\end{center}
\caption{ Above we give the complete list of simple Euclidean Jordan algebras and their rotation ( automorphism), Lorentz ( reduced structure) and Conformal ( linear fractional) groups.}
The symbols ${\mathbb R}$, ${\mathbb C}$, ${\mathbb H}$, ${\mathbb O}$
represent the four division algebras.  For the Jordan algebras
$J_n^{\mathbb A}$ of $n\times n$ hermitian matrices over $\mathbb{A}$ the norm form is the determinental form (or its
generalization to the quaternionic and octonionic matrices). \label{euclideanjordanlist}
\end{table}
Note that for Euclidean Jordan algebras $\Gamma_{(1,d)}$ the automorphism,
reduced structure and M\"{o}bius groups are simply the rotation,
Lorentz and conformal groups of $(d+1)$-dimensional Minkowski
spacetime.There exist the following special isomorphisms between the
Jordan algebras of $2\times 2$ Hermitian matrices over the four
division algebras and the Jordan algebras of gamma matrices:

\begin{equation}
~\\
J_{2}^{\mathbb R} \simeq \Gamma_{(1,2)}~~~~;~~~~J_{2}^{\mathbb C} \simeq \Gamma_{(1,3)} \\
~~~~;~~~~
J_{2}^{\mathbb H} \simeq \Gamma_{(1,5)}~~~~;~~~~J_{2}^{\mathbb O} \simeq \Gamma_{(1,9)} \\
~\\
\end{equation}
The Minkowski spacetimes they correspond to are precisely the critical
dimensions for the existence of super Yang-Mills theories as well as
of the classical Green-Schwarz superstrings. These Jordan algebras are
all quadratic and their norm forms are precisely the quadratic
invariants constructed using the Minkowski metric. The spacetimes defined by simple Jordan algebras of degree three can be interpreted as extensions of Minkowskian spacetimes in critical dimensions by bosonic spinorial coordinates plus a dilaton and the adjoint identity implies the Fierz identities for the existence of the corresponding supersymmetric theories \cite{Sierra:1986dx,Gunaydin:2005zz}.

We should note two important  facts about Table \ref{euclideanjordanlist}. First, the
conformal groups of generalized space-times defined by Euclidean
(formally real) Jordan algebras all admit positive energy unitary
representations.  Hence they can be
given a causal structure with a unitary time evolution as in four
dimensional Minkowski space-time \cite{Gunaydin:1999jb,Mack:2004pv}. Second is the fact that the maximal
compact subgroups of the generalized conformal groups of formally real
Jordan algebras are simply the compact real forms of their structure groups
(which are the products of their generalized Lorentz groups with
dilatations).

 Conformal group $Conf(J)$  of a Jordan algebra $J$ is
generated by translations $T_{\mathbf{a}}$ , special conformal generators $K_{\mathbf{a}}$ , dilatations and Lorentz transformations  $M_{\mathbf{a}\mathbf{b}}$ ( $\mathbf{a,b} \in J$).
Lorentz transformations and dilatations generate  the structure algebra $\mathfrak{str}(J)$  of $J$
\cite{Gunaydin:1975mp,Gunaydin:1989dq,Gunaydin:1992zh}.  Lie algebra $\mathfrak{conf}(J)$ of the conformal group $Conf(J)$ has  a 3-grading
with respect to the generator $D$ of dilatations:
\begin{equation}
  \mathfrak{conf}(J) = K_{\mathbf{a}} \oplus M_{\mathbf{a}\mathbf{b}} \oplus T_{\mathbf{b}}
\end{equation}
Action of $\mathfrak{conf}(J)$  on the elements $\mathbf{x}$ of a Jordan algebra $J$   are  as follows \cite{Gunaydin:1992zh}:
\begin{eqnarray}
T_{\mathbf{a}} \mathbf{x}  = \mathbf{a} \nonumber \\
M_{\mathbf{a}\mathbf{b}} \mathbf{x} = \{ \mathbf{a}\mathbf{b}\mathbf{x} \} \\
K_{\mathbf{a}} \mathbf{x} =-\frac{1}{2} \{\mathbf{x}\mathbf{a}\mathbf{x}\} \nonumber
\end{eqnarray}
where $\{\mathbf{a}\mathbf{b}\mathbf{x}\}$ is the Jordan triple product
\begin{equation*}
  \{\mathbf{a}\mathbf{b}\mathbf{x}\}:= \mathbf{a}\circ (\mathbf{b}\circ \mathbf{x}) - \mathbf{b}\circ (\mathbf{a} \cdot \mathbf{x}) + (\mathbf{a}\circ \mathbf{b}) \circ \mathbf{x}
\end{equation*}
\begin{equation*}
  \mathbf{a},\mathbf{b},\mathbf{x} \in  J
\end{equation*}
with $\circ $ denoting the Jordan product. They satisfy the commutation relations
\begin{eqnarray}
[T_\mathbf{a}, K_\mathbf{b}]&= & M_{\mathbf{a}\mathbf{b}} \\ \nn
[M_{\mathbf{a}\mathbf{b}},T_\mathbf{c}]&=&T_{\{\mathbf{a}\mathbf{b}\mathbf{c}\}} \\ \nn
[M_{\mathbf{a}\mathbf{b}},K_\mathbf{c}] &=& K_{\{\mathbf{b}\mathbf{a}\mathbf{c}\}} \\ \nn
[M_{\mathbf{a}\mathbf{b}},M_{\mathbf{c}\mathbf{d}}]&=& M_{\{\mathbf{a}\mathbf{b}\mathbf{c}\}\mathbf{d}}- M_{\{\mathbf{b}\mathbf{a}\mathbf{d}\}\mathbf{c}} \nn
\end{eqnarray}
corresponding to the well-known Tits-Kantor-Koecher construction of Lie algebras from Jordan triple systems \cite{MR0321986,MR0146231,MR0228554}.
The generators  $M_{\mathbf{a}\mathbf{b}}$ can be decomposed as
\begin{equation}
  M_{\mathbf{a}\mathbf{b}}= D_{\mathbf{a},\mathbf{b}} + L_{\mathbf{a}\cdot \mathbf{b}}
\end{equation}
where $D_{\mathbf{a},\mathbf{b}}$ are the derivations that generate the automorphism (rotation) group of $J$
\begin{equation*}
  D_{\mathbf{a},\mathbf{b}} \mathbf{x} = \mathbf{a}\circ
  (\mathbf{b}\circ \mathbf{x}) - \mathbf{b}\circ (\mathbf{a} \circ
  \mathbf{x})
\end{equation*}
and $L_{c}$ denotes multiplication by the element $c\in J$. The dilatation
generator $D$ is proportional to the multiplication operator by the
identity element of $J$.

Choosing a basis $e_I$ and a conjugate basis $\Tilde{e}^I$  of a  Jordan algebra $J$  transforming covariantly and  contravariantly, respectively, under the action of the Lorentz ( reduced structure) group of $J$ one can expand an element
$\mathbf{x} \in J$ as \[ \mathbf{x} = e_I q^I=\tilde{e}^I q_I \]
In this basis one can write the generators of $\mathfrak{conf}(J)$ as differential
operators acting on the ``coordinates'' $q^I$\cite{Gunaydin:1992zh}. These generators can be twisted
by a unitary character $\lambda$ and take on a simple and elegant form
\begin{eqnarray}
  T_I & = &\frac{\partial}{\partial q^I}  \nonumber \\
  R^I_J& = &- \Lambda^{IK}_{JL} q^L \frac{\partial}{\partial q^K}  - \lambda  \delta^I_J \\
  K^I & =& \frac{1}{2} \Lambda^{IK}_{JL}  q^J q^L  \frac{\partial}{\partial q^K} + \lambda q^I \nonumber \\
\end{eqnarray}
where
\begin{equation*}
   \Lambda_{KL}^{IJ} := \delta_K^I \delta_L^J + \delta_L^I \delta^J_K - \frac{4}{3} C^{IJM} C_{KLM}
\end{equation*}
They satisfy the commutation relations
\begin{eqnarray}
& [ T_I, K^J ] =& - R^J_I \\
&[ R^J_I , T_K ] = &\Lambda_{IK}^{JL} T_L \\
&[R^J_I , K^K ] = &- \Lambda_{IK}^{JL} K^L
\end{eqnarray}
The generator of the rotation (automorphism) subgroup are simply
\begin{equation}
A_{IJ} = R_I^J -R_J^I
\end{equation}

\section{U-duality Orbits of Extremal Black Hole Solutions of $5D$ Supergravity Theories with Symmetric Target Manifolds and their Spectrum Generating Conformal Extensions}
\renewcommand{\theequation}{\arabic{section}.\arabic{equation}}
Orbits of the spherically symmetric stationary BPS  black holes (BH) with non-vanishing entropy  under the action of U-duality groups of $N=2$ MESGT's
with symmetric target spaces were given in \cite{Ferrara:1997uz}.  In the same work the orbits with non-vanishing cubic invariants that are non-BPS were also classified. These latter orbits describe extremal non-BPS black holes and corresponding  solutions to the  attractor equations were given in \cite{Ferrara:2006xx}.  In this section we shall review the solutions to the attractor equations in $5D$ MESGTs for extremal black holes , BPS as well as non-BPS, following \cite{Ferrara:2006xx}.

 Let us denote the  $(n+1)$
dimensional charge vector in an extremal BH background as $q_I$.  It is   given by  \eq q_I = \int_{S^3} H_I = \int_{S^3}
\stackrel{\circ}{a}_{IJ} *F^J \hspace{1cm} (I=0,1,...n) \en
 The black hole potential \cite{Ferrara:1996um,Ferrara:1997tw} that determines the attractor flow takes on the following form for $N=2$ MESGTs:
\begin{equation}
V(\phi, q) = q_I {\stackrel{\circ}{a}}^{IJ} q_J
\end{equation}
where ${\stackrel{\circ}{a}}^{IJ}$ is the inverse of the metric
${\stackrel{\circ}{a}}_{IJ}$ of the kinetic energy term of the
vector fields.
In terms of the central charge function
 \[ Z= q_Ih^I \]
 the
potential can be written as \eq V(q,\phi) = Z^2 + \frac{3}{2} g^{xy} \partial_x Z
\partial_y Z \label{eq:N2} \en
where \[ \partial_x Z = q_I h^I_{,x} = \sqrt{\frac{2}{3}} h^I_x \]
The critical points of the potential are determined by
the equation \eq \partial_x V= 2 ( 2Z\partial_x Z - \sqrt{3/2}
T_{xyz} g^{yy'}g^{zz'} \partial_{y'}Z \partial_{z'} Z ) =0
\label{eq:derpotential}\en
The BPS attractors are given by the solutions satisfying \cite{Ferrara:2006em,Chamseddine:1996pi}
  \eq
\partial_x Z=0 \en
at the critical points.
The non-BPS attractors are given  by  non-trivial solutions \cite{Ferrara:2006xx}
\eq 2 Z \partial_x Z =
\sqrt{\frac{3}{2}} T_{xyz} \partial^y Z \partial^z Z
\label{eq:nonbps} \en
 such that
\[ \partial^x Z \equiv g^{xx'} \partial_{x'} Z \neq 0 \]
at the critical points.
The equation \ref{eq:nonbps} can be inverted using the relation
 \eq
q_I = h_I Z -\frac{3}{2} h_{I,x} \partial^x Z \en
For BPS attractors satisfying $\partial_x Z=0$ this gives \eq
q_I = h_I Z \en and for non-BPS attractors satisfying $\partial_x Z \neq 0$ we get \eq q_I = h_I Z
- (3/2)^{3/2} \frac{1}{2Z} h_{I, x} T^{xyz}
\partial_y Z \partial_z Z \en

Since the BPS attractor solution
with non-vanishing entropy   \cite{Ferrara:2006em,Chamseddine:1996pi} is given by  $ \partial_x
Z=0 $ , which is invariant under the automorphism group $Aut(J)$  of the underlying Jordan algebra $J$
, the orbits of BPS black hole solutions are of the form
\eq \mathcal{O}_{BPS} =  Str_0(J)/Aut(J) \en  and were listed in column 1 of table 1 of \cite{Ferrara:1997uz}  which we reproduce in Table \ref{5d_BPS_orbits}

\begin{table}
\begin{tabular}{|c|c|}  \hline
$J$ &$  \mathcal{O}_{BPS}= Str_0(J)/ Aut(J) $  \\ \hline $J_3^\mathbb{R}$ &
$
SL(3,\mathbb{R})/ SO(3) $ \\ \hline
$J_3^\mathbb{C}$   & $ SL(3,\mathbb{C})/ SU(3)  $ \\ \hline

$ J_3^\mathbb{H}$ & $ SU^*(6) / USp(6)  $ \\ \hline

$J_3^\mathbb{O}$  & $ E_{6(-26)} /F_4  $  \\ \hline
$\mathbb{R}\oplus \Gamma_{(1,n-1)}$ & $ SO(n-1,1)\times
SO(1,1) / SO(n-1) $ \\ \hline
\end{tabular}
\caption{ \label{5d_BPS_orbits} Above we list the orbits of spherically symmetric stationary BPS black hole solutions in $5D$  MESGTs defined by Euclidean Jordan algebras $J$  of degree three. U-duality and stability groups are given by the Lorentz  ( reduced structure ) and rotation ( automorphism) groups of $J$. }
\end{table}

The orbits for extremal non-BPS black holes with non-vanishing entropy are of the form \eq
\mathcal{O}_{non-BPS} = Str_0(J)/ Aut(J_{(1,2)}) \en where $ Aut(J_{(1,2)})$ is a noncompact real form of the automorphism group of $J$ and were  listed in column 2 of table
1 of \cite{Ferrara:1997uz} , which we reproduce in Table \ref{5d_orbits_nonBPS}.

\begin{table}
\begin{tabular}{|c|c|c|}  \hline
$ J$ & $ \mathcal{O}_{non-BPS}= Str_0(J)/ Aut(J_{(1,2)})  $& $ K \subset  Aut(J_{(1,2)})$ \\ \hline $J_3^{\mathbb{R}}$ & $ SL(3,\mathbb{R}) / SO(2,1) $  & SO(2)\\ \hline
$J_3^{\mathbb{C}}$ & $SL(3,\mathbb{C}) / SU(2,1)$ &  $SU(2)\times U(1)$\\ \hline
$J_3^{\mathbb{H}}$ & $SU^*(6) / USp(4,2)$ &  $USp(4)\times USp(2)$\\ \hline
$J_3^{\mathbb{O}}$ & $ E_{6(-26)} / F_{4(-20)}$ &  $SO(9)$ \\
 \hline
 $\mathbb{R}\oplus
\Gamma_{(1,n-1)}$
& $ SO(n-1,1) \times SO(1,1) / SO(n-2,1) $ &   $SO(n-2)$ \\ \hline
\end{tabular}
\caption{ \label{5d_orbits_nonBPS} Above we list  the
 orbits of non-BPS extremal black holes of  $N=2$ MESGT's with non-vanishing entropy in
$d=5$. The first column lists the Jordan algebras of degree 3 that
define these theories. The third column lists the maximal compact
subgroups $K$ of the stability group $ Aut(J_{(1,2)}) $.}
\end{table}

The entropy $S$ of an extremal black hole solution of $N=2$ MESGT  with charges $q_I$ is determined by the value of the black hole potential $V$ at the attractor  points
\eq
S = \left( V_{critical} \right)^{3/4}
\en

For$N=2$  MESGTs defined by Jordan algebras of degree 3, the tensor
$C_{IJK}$ is an invariant tensor of the U-duality group $Str_0(J)$.  Similarly the tensor $T_{abc}$ with " flat" indices
\[ T_{abc} = e_a^x e_b^y e_c^z T_{xyz} \] where $ e_a^x$ is the $n$-bein on the $n$-dimensional scalar manifold with metric $g_{xy}$ is
an invariant tensor of the maximal compact subgroup $Aut(J)$ of $Str_0(J)$ . In terms of flat indices the attractor equation becomes: \eq 2Z \partial_a Z =
\sqrt{3/2} T_{abc} \partial^b Z \partial^c Z \label{eq:attractor}
\en
Thus for  BPS attractor solution  $\partial_a Z=0$ one finds
 \eq
S_{BPS} = \left(V_{BPS} \right)^{3/4} =Z_{BPS}^{3/2}  \en
 For extremal non-BPS attractors  $\partial_aZ \neq 0
$ , squaring the criticality condition  one finds  \eq 4Z^2
\partial_a Z
\partial_a Z = \frac{3}{2} T_{abc} T_{ab'c'} \partial_b Z
\partial_c Z \partial_b' Z \partial_c' Z \en
Then using the identity
\[ T_{a(bc} T^a_{b'c')}= \frac{1}{2} g_{(bc} g_{b'c')} \]
valid only for  MESGTs defined by Jordan algebras of degree three
one obtains  \eq \partial_a Z \partial_a Z = \frac{16}{3} Z^2 \en
 Hence the entropy of extremal  non-BPS black holes are given by\footnote{We should stress that both  for  BPS as well as extremal non-BPS black holes the quantities appearing in the above formulas are evaluated at the corresponding attractor points.}

 \eq
 S_{non-BPS}= V_{non-BPS}^{4/3} = \left( Z^2 + \frac{3}{2} \partial_a Z \partial_a Z \right)^{3/4} = (3 Z_{non-BPS} )^{3/2}
 \label{eq:special}
 \en

 By differentiating equation \ref{eq:derpotential}  one  finds a general
expression for the Hessian of the black hole potential around the critical points \eqn \frac{1}{4} D_x
\partial_y V &=& \frac{2}{3} g_{xy} Z^2 + \partial_x Z \partial_y Z
-2\sqrt{\frac{2}{3}} T_{xyz} g^{zw} \partial_w Z Z \label{eq:masses} \\
&& +
T_{xpq}T_{yzs} g^{pz} g^{qq'}g^{ss'}\partial_{q'} Z \partial_{s'} Z \nn  \\
&=& \frac{2}{3}( g_{xz} Z -\sqrt{\frac{3}{2}} T_{xzp} \partial^p Z )
( g_{yz}Z - \sqrt{\frac{3}{2}} T_{yzq} \partial^q Z ) + \partial_x Z
\partial_y Z  \nn \enn
Thus for  BPS critical
points for which we have  $\partial_xZ=0$  the Hessian is given simply as \eq \partial_x \partial_y V =\frac{8}{3}
g_{xy} Z^2 \label{eq:positivemass} \en which is the same result as
in $d=4$ \cite{Ferrara:1997tw}. Since the metric of the scalar manifold is positive definite  the above formula
implies that the scalar fluctuations have positive square mass reflecting the
the attractor nature of the BPS critical points.

For non-BPS extremal critical points of the black hole potential  the Hessian has flat directions and is  positive semi-definite \cite{Ferrara:2006xx}.

The orbits of BPS black hole solutions of $N=8$ supergravity theory in five dimensions were also given in \cite{Ferrara:1997uz}. The $1/8$ BPS black holes with non-vanishing entropy has the orbit
\eq
\mathcal{O}_{1/8-BPS} = \frac{E_{6(6)}}{F_{4(4)}} =\frac{Str_0(J_3^{\mathbb{O}_S})}{Aut(J_3^{\mathbb{O}_S}) }\en
where $\mathbb{O}_S$ stands for the split octonions and $J_3^{\mathbb{O}_S}$ is the split exceptional Jordan algebra.
Note that in contrast to the exceptional $N=2$ MESGT theory defined by the real exceptional Jordan algebra that has two different orbits with nonvanishing entropy the maximal supergravity has only one such orbit. On the other hand maximal supergravity theory admits 1/4 and 1/2 BPS black hole solutions with vanishing entropy \cite{Ferrara:1997ci}. Their orbits under U-duality are \cite{Ferrara:1997uz}
\eq
\mathcal{O}_{1/4-BPS} = \frac{E_{6(6)}}{O(5,4)\circledS T_{16}} \en
\eq  \mathcal{O}_{1/2-BPS} = \frac{E_{6(6)}}{O(5,5)\circledS T_{16}} \en
where $\circledS$ denotes the semi-direct product and $T_{16}$ is the group of translations transforming in the spinor (16) of $SO(5,5)$.
Vanishing entropy means vanishing cubic norm. Thus the  black hole solutions corresponding to vanishing entropy has additional symmetries beyond the five dimensional U-duality group, namely they are invariant under the generalized special conformal transformations of the underlying Jordan algebras. This is complete parallel to the invariance of light-like vectors under special conformal transformations in four dimensional Minkowski spacetime which can be coordinatized by the elements of the Jordan algebra $J_2^{\mathbb{C}}$. Acting on a black hole solution with non-vanishing entropy these special conformal transformations change their norms and hence the entropy. Hence the conformal groups of  Jordan algebras were proposed as spectrum generating symmetry groups of black hole solutions of MESGTs defined by them \cite{Ferrara:1997uz,Gunaydin:2000xr,Gunaydin:2005gd,Gunaydin:2004ku}. Since the conformal groups of Jordan algebras of degree three are isomorphic to the U-duality groups of the corresponding four dimensional supergravity theories obtained by dimensional reduction this implies that the four dimensional U-duality groups must act as spectrum generating symmetry groups of the corresponding five dimensional theories.
Since it was first made, there have been several works  relating black hole solutions in four and
five dimensions ( $4D/5D$  lift)
\cite{Gaiotto:2005xt,Gaiotto:2005gf,Elvang:2005sa,Pioline:2005vi} that lend support to the proposal
that four dimensional U-duality
groups act as spectrum generating conformal symmetry groups of five
dimensional supergravity theories from which they
descend.

\section{4D, $N=2$  Maxwell-Einstein Supergravity Theories with Symmetric Target Spaces and Freudenthal Triple Systems}
\setcounter{equation}{0}
Under  dimensional reduction on a torus  $5D$, $N=2$
MESGTs with $(n_V-1)$ vector multiplets lead to $4D$, $N=2$ MESGTs  with $n_V$ vector multiplets, with the extra vector multiplet coming from the $5D$ graviton supermultiplet.  The metric of the target space of the four-dimensional
scalar fields of dimensionally reduced theories were first given in
\cite{Gunaydin:1983bi} in the so-called unbounded realization of their geometries. More precisely, the resulting  four
dimensional target spaces are generalized upper half-spaces (tube
domains) over the convex cones defined by the cubic norm. They are parameterized by
complex coordinates \cite{Gunaydin:1983bi},
\begin{equation}
   z^{I}:=\frac{1}{\sqrt{3}}A^{I} + \frac{i}{\sqrt{2}}\tilde{h}^{I}
\end{equation}
where $A^{I}$ denote the $4D$ scalars descending from the $5D$ vectors.   Imaginary components of $z^I$ are given by
\begin{equation}
   \tilde{h}^{I}:=e^\sigma h^{I}. \label{htildeI}
\end{equation}
where $\sigma$ is the scalar field ( dilaton ) coming from $5D$ graviton and $h^I$ were defined above. They
satisfy the positivity condition
\begin{equation*}
  \mathcal{V}(\tilde{h}^I) = C_{IJK} \tilde{h}^{I}\tilde{h}^{J}\tilde{h}^{K}= e^{3\sigma} >0
\end{equation*}

Geometry of four dimensional $N=2$ MESGTs obtained by dimensional
reduction from five dimensions (R-map) was later referred to as  ``very special
geometry'' and  studied extensively \footnote{ See for example  \cite{deWit:1991nm} and the references therein.}.
The full bosonic sector of $4D$ theories obtained by dimensional reduction from gauged $5D$, $N=2$ Yang-Mills Einstein supergravity coupled to tensor
multiplets and their reformulation in the standard language of special K\"ahler
geometry was given in \cite{Gunaydin:2005bf}, which we follow in our
summary here, restricting ourselves to the ungauged MESGT theory without tensors.

As is well-known one can interpret the $n_V$ complex coordinates $z^{I}$ of dimensionally reduced MESGTs   as
inhomogeneous coordinates of a $(n_V+1)$-dimensional complex vector with coordinates $X^A$
\begin{equation}
 X^A=  \left( \begin{array}{c} X^{0}\\ X^{I} \end{array} \right) =
       \left( \begin{array}{c}    1 \\ z^{I} \end{array} \right) .
\end{equation}
where the capital Latin indices $A,B,C,\cdots $ run from 0 to $n_V$.
Taking as ``prepotential'' the cubic form defined by the C-tensor coming from 5 dimensions
\begin{equation}
  F(X^A)=-\frac{\sqrt{2}}{3} C_{IJK}\frac{X^{I}X^{J}X^{K}}{X^{0}}
\label{prepot}
\end{equation}
and using the symplectic section
\begin{equation}
  \left( \begin{array}{c} X^A\\ F_{A}   \end{array} \right)  =
  \left( \begin{array}{c} X^A\\ \partial_{A}F \end{array} \right) \equiv
 \left( \begin{array}{c}  X^A\\ \frac{\partial F}{\partial X^{A}} \end{array}
\right)
\end{equation}
one gets  the K\"{a}hler potential
\begin{eqnarray}\label{kahlerpotential}
\mathcal{K}(X,\bar{X})&:=&-\ln [i\bar{X}^{A}F_{A}-iX^{A}\bar{F}_{A}] \label{symK}\\ \nonumber
&=&-\ln  \left[  i\frac{\sqrt{2}}{3}C_{IJK}(z^{I}-\bar{z}^{I})(z^{J}-\bar{z}^{J})(z^{K}-\bar{z}^{K})    \right] 
\end{eqnarray}
which agrees precisely with the K\"ahler potential obtained in \cite{Gunaydin:1983bi}.
The  ``period matrix'' that determines the kinetic terms of the vector fields in four dimensions is given by
\begin{equation}
\mathcal{N}_{AB}:=\bar{F}_{AB}+2i\frac{\textrm{Im}(F_{AC})\textrm{Im}(F_{BD})X^{C}X^{D}}{\textrm{Im}(F_{CD})X^{C}X^{D}}
\end{equation}
where $F_{AB}\equiv \partial_{A}\partial_{B}F$ etc.  Components of the resulting
period matrix $\mathcal{N}_{AB}$ under dimensional reduction are:
\begin{eqnarray}
\mathcal{N}_{00}&=&-\frac{2\sqrt{2}}{9\sqrt{3}}C_{IJK}A^{I}A^{J}A^{K}
-\frac{i}{3} \left( e^{\sigma}{\stackrel{\circ}{a}}_{IJ} A^{I}A^{J}+\frac{1}{2} e^{3\sigma} \right)\\
\mathcal{N}_{0I}&=&\frac{\sqrt{2}}{3}C_{IJK}A^{J}A^{K}+\frac{i}{\sqrt{3}} e^{\sigma} {\stackrel{\circ}{a}}_{IJ} A^{J}\\
\mathcal{N}_{IJ}&=&  -\frac{2\sqrt{2}}{\sqrt{3}}C_{IJK}A^{K}-ie^{\sigma}
{\stackrel{\circ}{a}}_{IJ} \label{NIJ}
\end{eqnarray}
The prepotential (\ref{prepot}) leads to the K\"ahler metric
\begin{equation}\label{metric}
  g_{I\bar{J}}  \equiv \partial_{I}\partial_{\bar{J}}\mathcal{K} = \frac{3}{2} e^{-2\sigma} {\stackrel{\circ}{a}}_{IJ}
\end{equation}
for the scalar manifold $\M_{4}$ of four-dimensional theory, where ${\stackrel{\circ}{a}}_{IJ}$ is the "metric" of the kinetic energy term of the vector fields of the $5D$ theory.  Above we denoted  the field strength of the vector field that comes from the
graviton in five dimensions as
$F_{\mu\nu}^{0}$.
The  bosonic sector of dimensionally reduced Lagrangian can then be written as
\begin{eqnarray}
   e^{-1}\mathcal{L}^{(4)} &=&-\frac{1}{2}R  -
   g_{I\bar{J}}    (\partial_{\mu}z^{I})(\partial^{\mu} \bar{z}^{J})
   \nonumber \\
   &&+\frac{1}{4}\textrm{Im}(\mathcal{N}_{AB})F_{\mu\nu}^{A}F^{\mu\nu B}-\frac{1}{8}
   \textrm{Re} (\mathcal{N}_{AB})\epsilon^{\mu\nu\rho\sigma}
   F_{\mu\nu}^{A}F_{\rho\sigma}^{B}.\label{redlag1b}
\end{eqnarray}

Since the complex scalar fields $z^{I}$ of the four dimensional theory are restricted to the domain $
\mathcal{V}(Im(z)) > 0$, the scalar manifolds of 4D, $N=2$ MESGT's
defined by Euclidean Jordan algebras $J$ of degree three are simply
the K\"ocher ``upper half spaces'' of the corresponding  Jordan algebras, which
belong to the family of Siegel domains of the first kind
\cite{MR1718170}.  The ``upper half spaces'' of Jordan algebras can be
mapped into bounded symmetric domains, which can be realized as
hermitian symmetric spaces of the form
\begin{equation}
  \mathcal{M}_4= \frac{Conf(J)}{\widetilde{Str}{J}}
\end{equation}
where $Conf(J)$ is the conformal group of the Jordan algebra $J$ and its maximal compact subgroup $\widetilde{Str}{J}$ is the compact real form of its structure group $Str(J)$.    The K\"ahler potential
\ref{kahlerpotential} that one obtains directly under dimensional
reduction from 5 dimensions is given by the ``cubic light-cone''
\begin{equation}
  \mathcal{V}(z-\bar{z}) =C_{IJK}(z^{I}-\bar{z}^{I})(z^{J}-\bar{z}^{J})(z^{K}-\bar{z}^{K})
\end{equation}
which is manifestly invariant under the 5 dimensional U-duality group
$Str_0(J)$ and real translations
\begin{equation*}
 Re (z^I)\Rightarrow Re(z^I) + a^I
\end{equation*}
\begin{equation*}
  a^I \in \mathbb{R}
\end{equation*}
which follows from Abelian gauge invariances of vector fields of the
five dimensional theory. Under dilatations it gets simply
rescaled. Infinitesimal action of special conformal generators $K^{I}$ of $Conf(J)$
on the ``cubic light-cone'' yields  \cite{Gunaydin:2000xr}
\begin{equation}
   K^{I} \mathcal{V}(z-\bar{z}) = (z^{I} + \bar{z}^{I}) \mathcal{V}(z-\bar{z})
\end{equation}
which can be integrated to give the global transformation of the form
\begin{equation}
\mathcal{V}(z-\bar{z}) \Longrightarrow f(z^{I}) \bar{f}( \bar{z}^{I}) \mathcal{V}(z-\bar{z})
\end{equation}
This shows that the cubic light-cone defined by $\mathcal{V}(z-\bar{z}) =0$ is
invariant under the full conformal group $Conf(J)$. Furthermore, the
above global conformal group action  leaves the metric $g_{I\bar{J}}$ invariant
since it simply induces  a K\"ahler transformation of the
K\"ahler potential  $\ln \mathcal{V}(z-\bar{z})$.

In $N=2$ MESGTs defined by Euclidean Jordan algebras $J$ of degree
three, one-to-one correspondence between vector fields of five
dimensional theories (and hence their charges) and elements of $J$
gets extended, in four dimensions, to a one-to-one correspondence
between field strengths of vector fields {\it plus} their magnetic
duals and Freudenthal triple systems defined over $J$
\cite{Gunaydin:1983bi,Gunaydin:2000xr,Ferrara:1997uz,Gunaydin:2005gd,Gunaydin:2005zz}. An
element $X$ of Freudenthal triple system (FTS) $\mathcal{F}(J) $ \cite
{MR0170974,MR0063358} over $J$ can be represented formally as a
$2\times 2$ ``matrix'':
\begin{equation}X= \left(
\begin{array}{ccc}
\alpha &  & \mathbf{x} \\
&  &  \\
\mathbf{y} &  & \beta
\end{array}
\right) \in \mathcal{F}(J)
\end{equation}
where $\alpha$, $\beta \in \mathbb{R}$ and $\mathbf{x}$,$\mathbf{y} \in J$

Denoting the ``bare'' four dimensional graviphoton field strength and
its magnetic dual as $F_{\mu \nu }^{0}$ and
$\widetilde{F}_{0}^{\mu\nu}$, respectively, we have the correspondence
\begin{equation*}
\left(
  \begin{array}{ccc}
     F_{\mu \nu }^{0} &  & F_{\mu \nu }^{I} \\ &  &  \\
     \widetilde{F}_{I}^{\mu \nu } &  & \widetilde{F}_{0}^{\mu \nu }
  \end{array}
\right) \Longleftrightarrow
 \left(
    \begin{array}{ccc}
        e_0 &  & e_I \\
            &  &  \\
\tilde{e}^I &  & \tilde{e}^0
    \end{array}
 \right) \in \mathcal{F}(J),
\end{equation*}
where $e_I (\tilde{e}^I)$ are the basis elements of $J$ (its conjugate $\tilde{J}$ ).
Consequently, one can associate with a black hole solution with
electric and magnetic charges (fluxes) $\left(q_0, q_I,
p^{0},p^{I}\right)$ of the 4D MESGT defined by $J$ an element of the FTS
$\mathcal{F}\left( J \right) $
\begin{equation}
\left(
\begin{array}{ccc}
p^{0}e_0 &  & p^{I} e_I\\
&  &  \\
q_{I} \tilde{e}^I &  & q_{0}\tilde{e}^0
\end{array}
\right)
\in \mathcal{F}(J) ,
\end{equation}
U-duality group $G_4$ of such a four dimensional MESGT acts as 
automorphism group of the FTS $\mathcal{F}(J)$, which is endowed with an
invariant symmetric quartic form and a skew-symmetric bilinear
form. The entropy of an extremal black with charges $ (p^0, p^I,q_0 ,q_I) $ is determined by the quartic invariant $\mathcal{Q}_4 (q,p)$ of $\mathcal{F}(J)$. With this identification the orbits of extremal black holes of 4D, $N=2$ MESGT's with symmetric scalar manifolds were classified in \cite{Ferrara:1997uz,Bellucci:2006xz}.

Upon further dimensional reduction to three dimensions (C-map) $N=2$
MESGTs lead to $N=4$, $d=3$ quaternionic K\"ahler $\sigma$ models
coupled to supergravity \cite{Gunaydin:1983bi,Ferrara:1989ik}. In
Table \ref{scalarmanifolds}  we give the symmetry groups of $N=2$ MESGTs defined by Euclidean Jordan
algebras in $d=5,4$ and $3$ dimensions and their scalar manifolds. We
should note that five and three dimensional U-duality symmetry groups
$Str_0(J)$ and $QConf(J)$, respectively, act as symmetries of
supergravity Lagrangians, while four dimensional U-duality groups
$Conf(J)$ are on-shell symmetries.
\begin{table}
\begin{equation} \nonumber
  \begin{array}{|c|c|c|c|} \hline
  ~ &  \mathcal{M}_5 = &  \mathcal{M}_4 = &  \mathcal{M}_3 =  \\
  J & \mathop\mathrm{Str}_0 \left(J\right)/ \mathop\mathrm{Aut}\left(J\right) &
            \mathop\mathrm{Conf} \left(J\right) / \mathop\mathrm{\widetilde{Str}}\left(J\right) &
       \mathop\mathrm{QConf} (\mathcal{F}(J))/
                            \widetilde{ \mathop\mathrm{Conf}\left(J\right)} \times \mathrm{SU}(2) \\[7pt]
        \hline
       J_3^\mathbb{R} & \mathrm{SL}(3, \mathbb{R}) / \mathrm{SO}(3) &
                        \mathrm{Sp}(6, \mathbb{R}) / \mathrm{U}(3)  &
            \mathrm{F}_{4(4)} / \mathrm{USp}(6) \times \mathrm{SU}(2)  \\ [7pt]
       J_3^\mathbb{C} & \mathrm{SL}(3, \mathbb{C}) / \mathrm{SU}(3) &
                        \mathrm{SU}(3, 3) / \mathrm{S}\left(\mathrm{U}(3) \times \mathrm{U}(3)\right) &
            \mathrm{E}_{6(2)} / \mathrm{SU}(6) \times \mathrm{SU}(2)   \\ [7pt]
       J_3^\mathbb{H} & \mathrm{SU}^\ast(6) / \mathrm{USp}(6) &
                        \mathrm{SO}^\ast(12) / \mathrm{U}(6) &
            \mathrm{E}_{7(-5)} / \mathrm{SO}(12) \times \mathrm{SU}(2)  \\[7pt]
       J_3^\mathbb{O} & \mathrm{E}_{6(-26)} / \mathrm{F}_4 &
                        \mathrm{E}_{7(-25)} / \mathrm{E}_6 \times \mathrm{U}(1) &
            \mathrm{E}_{8(-24)} / \mathrm{E}_7 \times \mathrm{SU}(2)  \\[7pt]
       \mathbb{R} \oplus \Gamma_{(1,n-1)} &
                      \frac{  \mathrm{SO}(n-1,1) \times \mathrm{SO}(1,1)}{ \mathrm{SO}(n-1)} &
                       \frac{ \mathrm{SO}(n,2) \times \mathrm{SU}(1,1) }{
                             \mathrm{SO}(n) \times \mathrm{SO}(2) \times \mathrm{U}(1)} &
           \frac{ \mathrm{SO}(n+2,4)}{ \mathrm{SO}(n+2) \times \mathrm{SO}(4)} \\ [7pt]
            \hline
  \end{array}
\end{equation}
\caption{\label{scalarmanifolds}
Above we list the scalar manifolds $\mathcal{M}_d$ of $N=2$ MESGT's
defined by Euclidean Jordan algebras $J$ of degree 3 in $d=3,4,5$
dimensions. $J_3^\mathbb{A}$ denotes the Jordan algebra of $3 \times 3$
Hermitian matrices over the division algebra $\mathbb{A} = \mathbb{R}$,
$\mathbb{C}$, $\mathbb{H}$, $\mathbb{O}$. The last row
$\mathbb{R} \oplus \Gamma_{(1,n-1)}$ are the reducible Jordan
algebras which are direct sums of Jordan algebras $\Gamma_{(1,n-1)}$ defined
by a quadratic form $\mathbb{Q}$ of Minkowskian signature and  one dimensional Jordan algebra
$\mathbb{R}$. $\mathop\mathrm{\widetilde{Str}}\left(J\right)$ and
$\widetilde{\mathop\mathrm{Conf}}\left(J\right)$ denote the compact
real forms of the structure group $\mathop\mathrm{Str}\left(J\right)$
and conformal group $\mathop\mathrm{Conf}\left(J\right)$ of a Jordan
algebra
$J$. $\mathop\mathrm{QConf}\left(\mathcal{F}\left(J\right)\right)$
denotes the quasiconformal group defined by the FTS
$\mathcal{F}\left(J\right)$ defined over $J$.
}
\end{table}

\section{U-duality Orbits of Extremal Black Holes of  $4D$, $N=2$ MESGTs with Symmetric Scalar Manifolds and of $N=8$ Supergravity and their Spectrum Generating Quasiconformal Extensions }
\setcounter{equation}{0}
The discussion of the orbits  extremal black holes of extended supergravity theories with symmetric scalar manifolds were covered in Sergio Ferrara's lectures \cite{Bellucci:2009qv}. Referring to Ferrara's lectures for details including the related recent developments I will briefly summarize the results for $N=2$ MESGTs and $N=8$ supergravity in this section following \cite{Bellucci:2006xz}.
As in the five dimensional case,  the $4D$ extremal
black hole attractor equations are simply  the criticality conditions for
the black hole scalar potential which can be written as \cite{Ceresole:1995ca,Ferrara:1996um}
\begin{equation}
V_{BH}\equiv \left| Z\right| ^{2}+G^{I\overline{J}}( D_{I}Z ) (\overline{D}_{
\overline{J}}\overline{Z} ) \label{VBH-def}
\end{equation}
where $Z$ is the central charge function.
The criticality condition is \cite{Ferrara:1997tw}:
\begin{equation}
\partial _{I}V_{BH}=0
\end{equation}
implies
\begin{equation}
2\overline{Z}D_{I}Z+iC_{IJK}G^{J 
\overline{J}}G^{K\overline{K}}\overline{D}_{\overline{J}}\overline{Z} 
\overline{D}_{\overline{K}}\overline{Z}=0
\label{criticality_4d}
\end{equation}
$C_{IJK}$ is the completely symmetric, covariantly holomorphic
tensor of special K\"{a}hler geometry that satisfies
\begin{equation}
\overline{D}_{\overline{L}}C_{IJK}=0,~~~D_{[L}C_{I]JK}=0
\end{equation}
where square brackets denote antisymmetrization.
For symmetric special K\"{a}hler manifolds the tensor $C_{IJK}$ is
covariantly constant:\textbf{\ }
\begin{equation}
D_{I}C_{JKL}=0
\end{equation}
which implies the four dimensional counterpart of the adjoint identity \cite{Gunaydin:1983bi,Cremmer:1984hc}
\begin{equation}
G^{K\overline{K}}G^{M\overline{J}}C_{M(PQ}C_{IJ)K}\overline{C}_{\overline{K} 
\overline{I}\overline{J}}=\frac{4}{3}C_{(IJP} G_{ Q) \overline{I}}
\end{equation}

The $\frac{1}{2}$-BPS attractors are given by the following
solution  of attractor equations \cite{Ferrara:1997tw}
\begin{equation}
Z\neq 0,\text{ }D_{I}Z=0 \,\,\,\, \forall  \,\,\, I=1,...,n_{V}.
\label{BPS}
\end{equation}
The orbits of the 1/2-BPS black hole solutions with positive quartic invariants  of $N=2$ MESGTs with symmetric target spaces were
given in \cite{Ferrara:1997uz} and are listed in column 1 of Table  \ref{4dOrbits}.
In \cite{Ferrara:1997uz} a second family of orbits with non-vanishing quartic invariants were also given. They correspond to non-BPS extremal black holes and the respective solutions to the attractor equations were given in \cite{Bellucci:2006xz}. In addition there  exist another family of non-BPS extremal black holes with non-vanishing quartic invariant and vanishing central charge \cite{Bellucci:2006xz}.
The complete list of orbits of BPS and extremal non-BPS black holes is given in Table \ref{4dOrbits}.

\begin{table}\label{4dOrbits}
\begin{center}
\begin{tabular}{|c|c|c|c|}
\hline
J & $
\begin{array}{c}
\\
\frac{1}{2}\text{-BPS orbits } \\
~~\mathcal{O}_{\frac{1}{2}-BPS} \\
~
\end{array}
$ & $
\begin{array}{c}
\\
\text{non-BPS, }Z\neq 0\text{ orbits} \\
\mathcal{O}_{non-BPS,Z\neq 0}~ \\
~
\end{array}
$ & $
\begin{array}{c}
\\
\text{non-BPS, }Z=0\text{ orbits} \\
\mathcal{O}_{non-BPS,Z=0}~ \\
~
\end{array}
$ \\ \hline\hline
$
\begin{array}{c}
\\
- \\
~
\end{array}
$ & $\frac{SU(1,n+1)}{SU(n+1)}~$ & $-$ & $\frac{SU(1,n+1)}{SU(1,n)}~$ \\
\hline
$
\begin{array}{c}
\\
\mathbb{R} \oplus \Gamma_{(1,n-1)}  \\
~
\end{array}
$ & $\frac{SU(1,1)\otimes SO(2,2+n)}{SO(2)\otimes SO(2+n)}~$ & $\frac{%
SU(1,1)\otimes SO(2,2+n)}{SO(1,1)\otimes SO(1,1+n)}~$ & $\frac{%
SU(1,1)\otimes SO(2,2+n)}{SO(2)\otimes SO(2,n)}$ \\ \hline
$
\begin{array}{c}
\\
 J_3^{\mathbb{O}} \\
~
\end{array}
$ & $\frac{E_{7(-25)}}{E_{6}}$ & $\frac{E_{7(-25)}}{E_{6(-26)}}$ & $\frac{%
E_{7(-25)}}{E_{6(-14)}}~$ \\ \hline
$
\begin{array}{c}
\\
 J_3^{\mathbb{H}}  \\
~
\end{array}
$ & $\frac{SO^{\ast }(12)}{SU(6)}~$ & $\frac{SO^{\ast }(12)}{SU^{\ast }(6)}~$
& $\frac{SO^{\ast }(12)}{SU(4,2)}~$ \\ \hline
$
\begin{array}{c}
\\
 J_3^{\mathbb{C}}  \\
~
\end{array}
$ & $\frac{SU(3,3)}{SU(3)\otimes SU(3)}$ & $\frac{SU(3,3)}{SL(3,\mathbb{C})}$
& $\frac{SU(3,3)}{SU(2,1)\otimes SU(1,2)}~$ \\ \hline
$
\begin{array}{c}
\\
 J_3^{\mathbb{R}} \\
~
\end{array}
$ & $\frac{Sp(6,\mathbb{R})}{SU(3)}$ & $\frac{Sp(6,\mathbb{R})}{SL(3,\mathbb{%
R})}$ & $\frac{Sp(6,\mathbb{R})}{SU(2,1)}$ \\ \hline
\end{tabular}
\end{center}
\caption{ \label{4dOrbits} Non-degenerate orbits of $N=2$, $D=4$ MESGTs with symmetric scalar manifolds. Except for the first row all such theories originate from five dimensions and are defined by Jordan algebras that are indicated in the first column. }
\end{table}

The orbits of black hole solutions of $4D$ $N=8$ supergravity under the action of U-duality group
$E_{7(7)}$ were given in \cite{Ferrara:1997uz}.
 There exist two
classes of non-degenerate charge orbits of black hole solutions  with non-vanishing quartic
invariant $I_4$  constructed from the electric and magnetic charges transforming in $%
\mathbf{56}$ of $E_{7(7)}$ \cite{Ferrara:1997uz}. Depending on the sign of $I_4$
, one finds :
\begin{eqnarray}
I_{4} &>&0:\mathcal{O}_{\frac{1}{8}-BPS}=\frac{E_{7(7)}}{E_{6(2)}} \Longleftrightarrow \frac{1}{8}\text{-BPS;}   \\
&&  \notag \\
I_{4} &<&0:\mathcal{O}_{non-BPS}=\frac{E_{7(7)}}{E_{6(6)}}  \Longleftrightarrow \text{non-BPS.}
\end{eqnarray}

This is to be contrasted with the non-degenerate orbits of the exceptional $N=2$ supergravity with
U-duality group $E_{7(-25)}$, which has three non-degenerate orbits, one BPS and two non-BPS one of which has vanishing central charge.  On the other hand, in $N=8$ supergravity one has $1/4$ and $1/2$ BPS black holes with vanishing entropy \cite{Ferrara:1997ci}. The "light-like" orbits of these BPS black hole solutions with vanishing quartic invariant were given in \cite{Ferrara:1997uz}
There are 3 distinct cases depending on the number of
vanishing "eigenvalues" that lead to vanishing $\mathcal{Q}_4$. The
generic light-like orbit  for which a single eigenvalue
vanishes is
\begin{equation}
\frac{E_{7(7)}}{F_{4(4)}\circledS T_{26}}
\end{equation}
where $T_{26}$ is a 26 dimensional Abelian subgroup of $E_{7(7)}$ and $\circledS$ denotes semi-direct product.
The critical light-like orbit has two vanishing eigenvalues and
correspond to the 45 dimensional orbit
\begin{equation}
\frac{E_{7(7)}}{O(6,5)\circledS ( T_{32} \oplus T_1 )}
\end{equation}
The doubly critical light-like orbit with  three vanishing eigenvalues is
given by the 28 dimensional quotient space
\begin{equation}
\frac{E_{7(7)}}{E_{6(6)}\circledS T_{27}}
\end{equation}

As discussed above, four dimensional U-duality groups $G_4$ were
proposed as  spectrum generating conformal symmetry groups in five
dimensions that leave a cubic light-cone invariant. This raises the question, first investigated in \cite{Gunaydin:2000xr},  whether the
three dimensional U-duality groups $G_3$ could act as spectrum generating "conformal" groups of  corresponding four dimensional supergravity theories.  It is easy to show that there exist three dimensional U-duality groups that do not have any
conformal realizations in general. Some other three dimensional U-duality groups do not admit conformal realizations  on the $2n_V+2$ dimensional space of the FTS that defines the four dimensional theory.
However as was shown in \cite{Gunaydin:2000xr} the three dimensional
U-duality groups $G_3$ all have novel geometric realizations as quasi-conformal
groups on the vector spaces of FTS's  extended by an extra singlet coordinate that
leave invariant a generalized light-cone with respect to a quartic
distance function. The quasiconformal actions of three dimensional
U-duality groups $G_3$ were then  proposed as  spectrum generating symmetry
groups of corresponding four dimensional supergravity theories
\cite{Gunaydin:2000xr,Gunaydin:2004ku,Gunaydin:2003qm,Gunaydin:2005gd,Gunaydin:2005mx,Gunaydin:2007bg}.
We shall denote the quasiconformal groups defined over FTS's
$\mathcal{F}$ extended by a singlet coordinate as
$QConf(\mathcal{F})$. If the FTS is defined over a Jordan algebra $J$ of degree three we shall denote the corresponding quasiconformal
groups either as $QConf(\mathcal{F}(J))$ or simply as $QConf(J)$. The construction given in \cite{Gunaydin:2000xr} is covariant with respect to the automorphism group of the FTS, which is isomorphic to the $4D$ U-duality group of the corresponding supergravity. For
$N=2$ MESGTs defined by Jordan algebras of degree three,
quasiconformal group actions of their three dimensional U-duality
groups $G_3$ were given explicitly in \cite{Gunaydin:2005zz}, in a
basis covariant with respect to U-duality groups $G_6$ of corresponding six
dimensional supergravity theories.

\section{ Quasiconformal Realizations of Lie Groups and Freudenthal Triple Systems }
\setcounter{equation}{0}
In this section we shall review the general theory of quasiconformal
realizations of noncompact groups over Freudenthal triple systems that
was given in \cite{Gunaydin:2000xr}.

Every simple Lie algebra $\mathfrak{g}$ of dimension greater than three  can be given a 5-graded
decomposition\footnote{ This is to be contrasted with the three grading of generalized conformal groups. No real forms of exceptional Lie algebras $G_2 , F_4$ and $E_8$ admit such a three grading.}  , determined by one of its generators $\Delta$, such that
grade $\pm 2$ subspaces are one dimensional:
\begin{eqnarray}
\mathfrak{g} = \mathfrak{g}^{-2} \oplus \mathfrak{g}^{-1} \oplus
       \mathfrak{g}^0 \oplus \mathfrak{g}^{+1}
     \oplus \mathfrak{g}^{+2} \,.
\end{eqnarray}
where
\begin{equation}
\mathfrak{g}^0 = \mathfrak{h} \oplus \Delta
\end{equation}
and
\begin{equation}
[\Delta, \mathfrak{t}]= m \mathfrak{t} \; \;\;\; \forall \mathfrak{t} \in \mathfrak{g}^m \;\;,\; m=0,\pm1,\pm2
\end{equation}
Given such a 5-graded Lie algebra it can be
constructed over a Freudenthal triple system $\mathcal{F}$ which we shall denote as $ \mathfrak{g}(\mathcal{F}) $     \cite{MR0063358,kansko}. A
Freudenthal triple system (FTS) is defined as a vector space
$\mathcal{F}$ equipped with a triple product $(X,Y,Z)$
\begin{equation}
(X,Y,Z) \in \mathcal{F} \;\;\;\;\;\;\;\; \forall \;\; X,Y,Z \in \mathcal{F}
\end{equation}
that  satisfies the identities
\begin{eqnarray}
(X,Y,Z) &=&(Y,X,Z) +2\,\langle X,Y \rangle Z \,,\nonumber \\
(X,Y,Z) &=& (Z,Y,X) -2\,\langle X,Z \rangle Y \,,\nonumber \\
\langle (X,Y,Z), W \rangle  &=& \langle (X,W,Z),Y \rangle
                                -2\,\langle X, Z \rangle \langle Y ,W \rangle \,,\nonumber \\
(X,Y,(V,W,Z)) &=& (V,W,(X,Y,Z)
                             +((X,Y,V),W,Z) \nonumber \\
                          && {}+ (V,(Y,X,W),Z)  \,.\label{ftp56-rel}
\end{eqnarray}
and admits a skew symmetric bilinear
form \[ \langle X,Y\rangle = -\langle Y,X\rangle \in \mathbb{R} , \,\,\,\, \forall \,\, X,Y \in \mathcal{F}\] 
In the corresponding construction of
$\mathfrak{g}(\mathcal{F})$ one labels the generators belonging to
subspace $\mathfrak{g}^{+1}$ by the elements of $\mathcal{F}$
\begin{equation}
U_A \in \mathfrak{g}^{+1} \leftrightarrow \;\;\;\;\; A\in \mathcal{F}
\end{equation}
and through the  involution , that reverses the grading, elements of
$\mathfrak{g}^{-1}$ can also be labeled by elements of $\mathcal{F}$
\begin{equation}
   \tilde{U}_A \in \mathfrak{g}^{-1} \leftrightarrow \;\;\;\;\; A\in \mathcal{F}
\end{equation}
Elements of $\mathfrak{g}^{\pm1}$ generate the full Lie algebra
$\mathfrak{g}(\mathcal{F})$ by commutation. The
generators belonging to grade zero and grade $\pm 2$ subspaces are labelled by a pair of elements of $\mathcal{F}$
\begin{eqnarray}
 [U_A,\tilde{U}_B]         &\equiv & S_{AB} \;\; \in \mathfrak{g}^0            \\  \nn
 [U_A,U_B]            & \equiv & -K_{AB} \;\; \in \mathfrak{g}^2 \\  \nn
 [\tilde{U}_A,\tilde{U}_B] & \equiv & -\tilde{K}_{AB} \;\; \in \mathfrak{g}^{-2} \\ \nn
 \end{eqnarray}
Commutation relations of the generators of the Lie algebra $\mathfrak{g}$ can all be expressed in terms of the Freudenthal triple
product $(A,B,C)$
\begin{eqnarray}
 [S_{AB},U_C]            &=& -U_{(A,B,C)}          \\ \nn
 [S_{AB},\tilde{U}_C]    &=& - \tilde{U}_{(B,A,C)} \\ \nn
 [K_{AB},\tilde{U}_C] &=& U_{(A,C,B)} - U_{(B,C,A)} \\ \nn
 [\tilde{K}_{AB},U_C] &=& \tilde{U}_{(B,C,A)} -\tilde{U}_{(A,C,B)} \\ \nn
 [S_{AB},S_{CD}]   &=& -S_{(A,B,C)D}
                            -S_{C (B,A,D)} \\ \nn
 [S_{AB},K_{CD}]         &=&  K_{A\FTP{C}{B}{D}}
                              - K_{A\FTP{D}{B}{C}} \\ \nn]
 [S_{AB},\tilde{K}_{CD}] &=&  \tilde{K}_{\FTP{D}{A}{C}B}
                              - \tilde{K}_{\FTP{C}{A}{D}B}\\  \nn
 [K_{AB},\tilde{K}_{CD}] &=&   S_{\FTP{B}{C}{A}D}
                                -S_{\FTP{A}{C}{B}D}
                                -S_{\FTP{B}{D}{A}C}
                                +S_{\FTP{A}{D}{B}C} \\ \nn
\end{eqnarray}
Since the grade $\pm 2$ subspaces are one dimensional their generators can be written as
\begin{equation}
  K_{AB} :=K_{\langle A , B\rangle}:=\langle A , B\rangle  K
\end{equation}
\begin{equation}
  \tilde{K}_{AB} :=\tilde{K}_{\langle A , B\rangle}:= \langle A, B\rangle  \tilde{K}
\end{equation}
Now the defining identities of a FTS imply that
\begin{equation}
S_{AB}-S_{BA} = -2 \langle A, B\rangle \Delta
\end{equation}
where $\Delta$ is the generator that determines the 5-grading
\begin{eqnarray}
  [\Delta, U_A]&=&U_A \\ \nn
  [\Delta, \tilde{U}_A ]&=& - \tilde{U}_A \\ \nn
  [\Delta , K]& = &2 K \\ \nn
  [\Delta, \tilde{K} ]& =& - 2 \tilde{K}
\end{eqnarray}
and generates a distinguished  $sl(2)$ subalgebra together with $K, \tilde{K}$
\begin{equation}
  [K, \tilde{K}] = -2 \Delta
\end{equation}
The 5-grading of $\mathfrak{g}$ can then be recast  as
\begin{equation*}
   \mathfrak{g} = \tilde{K} \oplus \tilde{U}_A \oplus [S_{(AB)}+ \Delta ] \oplus U_A \oplus K
\end{equation*}
where
\begin{equation*}
  S_{(AB)} := \frac{1}{2} ( S_{AB} + S_{BA})
\end{equation*}
are the generators of the automorphism group $Aut(\mathcal{F})$ of $\mathcal{F}$ that  commute with $\Delta$
\begin{equation}
  [\Delta, S_{(AB)} ]=0
\end{equation}
The remaining  non-zero commutators are
\begin{eqnarray}
  [ U_A, \tilde{U}_B ] &=& S_{(AB)} - \langle A, B \rangle \Delta \\ \nn
  [ K, \tilde{U}_A]&=& -2 \tilde{U}_A \\ \nn
  [ \tilde{K}, U_A]&=&  2 \tilde{U}_A \\ \nn
  [ S_{(AB)}, K]&=&0 \\ \nn
\end{eqnarray}
Every FTS $\mathcal{F}$ admits a completely symmetric quadrilinear form which induces a quartic norm $\mathcal{Q}_4$. For an element $ X \in \mathcal{F}$ the quartic norm is
\begin{eqnarray}
\mathcal{Q}_4(X) := \frac{1}{48}\langle (X,X,X),X \rangle \label{e7-invariant}
\end{eqnarray}
which is invariant under the automorphism group $Aut(\mathcal{F})$ of
$\mathcal{F}$ generated by $S_{(AB)}$.

As was shown in \cite{Gunaydin:2000xr} one can realize the 5-graded
Lie algebra $\mathfrak{g}$ non-linearly as a quasiconformal Lie
algebra over a vector space $\mathcal{T}$ coordinatized by the
elements $X$ of the FTS $\mathcal{F}$ plus  an extra singlet variable $x$
\cite{Gunaydin:2000xr,Gunaydin:2005zz}:
\begin{equation}
\begin{split}
  \begin{aligned}
      K\left(X\right) &= 0 \\
      K\left(x\right) &= 2\,
  \end{aligned}
  & \quad
  \begin{aligned}
     U_A \left(X\right) &= A \\
     U_A\left(x\right) &= \left< A, X\right>
  \end{aligned}
   \quad
   \begin{aligned}
      S_{AB}\left(X\right) &= \left( A, B, X\right) \\
      S_{AB}\left(x\right) &= 2 \left< A, B\right> x
   \end{aligned}
 \\
 &\begin{aligned}
    \Tilde{U}_A\left(X\right) &= \frac{1}{2} \left(X, A, X\right) - A x \\
    \Tilde{U}_A\left(x\right) &= -\frac{1}{6} \left< \left(X, X, X\right), A \right> + \left< X, A\right> x
 \end{aligned}
 \\
 &\begin{aligned}
    \Tilde{K}\left(X\right) &= -\frac{1}{6} \,  \left(X,X,X\right) +  X x \\
    \Tilde{K}\left(x\right) &= \frac{1}{6} \,  \left< \left(X, X, X\right), X \right> + 2\,  \, x^2
 \end{aligned}
\end{split}
\end{equation}
The quasiconformal action of the Lie algebra $\mathfrak{g}(\mathcal{F})$  on the space $\mathcal{T}$ has a beautiful geometric interpretation. To see this one  defines  the quartic norm of a vector $\cX =(X,x)$ in the space $\mathcal{T}$ as
\begin{equation}
\cN_4(\cX) := \mathcal{Q}_4(X) - x^2
\end{equation}
where $\mathcal{Q}_4(X)$ is the quartic norm  of $X \in \mathcal{F}$ and
then a ``distance'' function between any two points $\cX=(X,x)$ and
$\cY=(Y,y) $ in $\mathcal{T}$ as
\begin{equation}
d(\cX,\cY):= \cN_4(\gd(\cX,\cY)
\end{equation}
where  $\gd(\cX,\cY)$ is the ``symplectic'' difference of two  vectors $\cX $ and $\cY$ :
\begin{equation}
  \gd(\cX,\cY) := (X-Y,x-y+\langle X, Y \rangle )= - \gd(\cY , \cX) 
\end{equation}
One can then show that the light-like separations with respect to this
quartic distance function
\begin{equation}
  d(\cX,\cY)=0
\end{equation}
is left invariant under  quasiconformal group action
\cite{Gunaydin:2000xr}. In other words quasiconformal groups are the invariance groups of "light-cones" defined by a quartic distance function.

\section{$3D$ U-duality Groups  as Spectrum Generating Quasiconformal Groups of $4D$ Supergravity Theories and Quantum Attractor Flows}
\renewcommand{\theequation}{\arabic{section}.\arabic{equation}}
\setcounter{equation}{0}
As explained above the vector field strengths plus their magnetic duals of a $4D$ supergravity  defined by a Jordan algebra $J$ of degree three are in one-to-one correspondence with the elements of the Freudenthal triple $\mathcal{F}(J)$ defined over $J$.
The automorphism group of  $\mathcal{F}(J)$
 is 
the  U-duality group $G_4$ of  the supergravity  defined by $J$ and is isomorphic to the conformal group $Conf(J)$ of $J$.
Furthermore, U-duality symmetry groups $G_3$ of the $3D$ supergravity theories they reduce to under dimensional reduction are the quasiconformal groups $QConf(J)$ of $\mathcal{F}(J)$. The U-duality groups of $N=2$ MESGTs defined by Jordan algebras of degree three in five, four and three dimensions are also the isometry groups  of their scalar manifolds in the respective dimensions. In  five dimensions scalar manifolds are 
\begin{equation*}
  \mathcal{M}_5 = \frac{Str_0(J)}{Aut(J)}
\end{equation*}
where $Str_0(J)$ and $Aut(J)$ are the reduced structure and
automorphism groups of  $J$, respectively. The 
scalar manifolds of these theories in four dimensions are
\begin{equation*}
 \mathcal{M}_4 = \frac{Conf(J)}{ \widetilde{Str}_0 (J) \times U(1)}
\end{equation*}
where $Conf(J)$ is the conformal group of the Jordan algebra $J$ and
$\widetilde{Str}_0(J)$ is the compact form of the reduced structure
group. Upon further dimensional reduction to three dimensions they
lead to scalar manifolds of the form
\begin{equation*}
 \mathcal{M}_3 = \frac{QConf(J)}{\widetilde{Conf}(J) \times SU(2)}
\end{equation*}
where $QConf(J)$ is the quasiconformal group associated with the
Jordan algebra $J$ and $\widetilde{Conf}(J)$ is the compact real form of
the conformal group of $J$. The complete list of the symmetric scalar manifolds in five , four and three dimensions are given in
Table \ref{scalarmanifolds}.

In the original proposal of \cite{Gunaydin:2000xr} that the three dimensional U-duality groups act as spectrum generating quasiconformal groups of the corresponding four dimensional supergravity theories 
the extra singlet coordinate that extends the 56 dimensional charge space ($p^A, q_A$ )  of black hole solutions of $N=8$ supergravity  was interpreted as the entropy $s$ of the black hole. The light cone condition on the 57 dimensional charge-entropy vector $(p^A,q_A,s)$ on which $G_3$ acts as a quasiconformal group then gives the well-established relation between  the entropy $s$ to  the quartic invariant $\mathcal{Q}_4$ constructed out of the charges 
\[ s^2 = \mathcal{Q}_4(p^A,q_A) \]

A concrete and precise implementation of the proposal that three
dimensional U-duality groups must act as spectrum generating quasiconformal
groups of  spherically
symmetric stationary BPS black holes of four dimensional supergravity
theories, was given  in
\cite{Gunaydin:2005mx,Gunaydin:2007bg,Gunaydin:2007qq}  which we will summarize in this section \footnote{ See also  \cite{Pioline:2008zz}.}. The basic starting point of these works is the fact that the
attractor equations \cite{Ferrara:1995ih,Ferrara:1996um} for a spherically symmetric stationary black hole of
four dimensional supergravity theories are equivalent to the equations for  geodesic
motion of a fiducial particle on the moduli space $\mathcal{M}^*_3$ of
the three dimensional supergravity obtained by reduction on a
time-like circle. The connection between the stationary black holes of $4d$ gravity coupled to matter and geodesic motion of a fiducial particle on the pseudo-Riemannian manifold $\mathcal{M}_3$  coupled to gravity in three dimensions  on a timelike circle   was first observed in
\cite{Breitenlohner:1987dg} \footnote{ This was  used in
\cite{Cvetic:1995kv,Cvetic:1995uj} to construct static and rotating
black holes in heterotic string theory.}.  More specifically, 
 a $4D$ supergravity theory with symmetric scalar manifold $\mathcal{M}_4 = G_4/K_4$ reduces on a space-like circle to a $3D$ supergravity with scalar manifold
\[ \mathcal{M}_3 = \frac{G_3}{K_3} \]
where $K_3$ is the maximal compact subgroup of $G_3$. The same theory dimensionally reduced on a time-like circle
leads to a theory with scalar manifold of the form \cite{Breitenlohner:1987dg}
\[ \mathcal{M}^* = \frac{G_3}{H_3} \]
where $H_3$ is a certain noncompact real form of $K_3$. 
Then the stationary, spherically symmetric solutions of
the four-dimensional equations of motion are equivalent to geodesic
trajectories on the three-dimensional scalar manifold
$\mathcal{M}_3^*=G_3/H_3$ \cite{Breitenlohner:1987dg}. For $N=2$ MESGTs  defined by Euclidean Jordan algebras of degree three
the resulting spaces $\mathcal{M}_3^*$ are para-quaternionic symmetric spaces of the form
\begin{equation}
\mathcal{M}_3^* = \frac{QConf(J)}{ Conf(J) \times SU(1,1)}
\end{equation}
where $Conf(J)$ is the isometry group of the scalar manifold $\mathcal{M}_4$ of the four dimensional theory.
In Table \ref{paraspaces} we reproduce a table from \cite{Gunaydin:2005mx} giving a complete list of  supergravity theories whose four dimensional isometry groups are conformal groups $Conf(J)$ of a Jordan algebra of degree three and the resulting scalar manifolds $\mathcal{M}_3^*$, which include all $N  \geq 4$ supergravity theories as well as $N=2$ MESGTs defined by Euclidean Jordan algebras. 

\begin{table}
\begin{tabular}{|c|c|c|c|c|} \hline
$n_Q$ & $n_V$ & $M_4$           & $\mathcal{M}_3^*$  & $J$  \\ \hline \hline
    8        & 1  & $\emptyset $  & $\frac{U(2,1)}{ U(1,1)\times U(1)}$ & $\mathbb{R}$                 \\ \hline
    8        & 2  & $\frac{SL(2,\mathbb{R})}{U(1)}$  & $\frac{G_{2,2}}{ SO(2,2)}$ & $\mathbb{R}$                 \\ \hline
    8        & 7  & $\frac{Sp(6,\mathbb{R})}{ SU(3) \times U(1)}$ & $\frac{F_{4(4)}}{ Sp(6,\mathbb{R}) \times SL(2,\mathbb{R})}$  &  $J_3^\mathbb{R}$ \\ \hline
    8        & 10 & $\frac{SU(3,3)}{ SU(3) \times SU(3) \times U(1)}$ & $\frac{E_{6(2)}}{
SU(3,3) \times SL(2,\mathbb{R})}$ &  $J_3^\mathbb{C}$ \\ \hline
    8        & 16 & $\frac{SO^*(12)}{ SU(6) \times U(1)}$ & $\frac{E_{7(-5)}}{
SO^*(12) \times SL(2,\mathbb{R})}$ &  $J_3^{\mathbb{H}}$ \\ \hline
    8        & 28 & $\frac{E_{7(-25)}}{E_{6} \times U(1)}$ & $\frac{E_{8(-24)}}{ E_{7(-25)} \times SL(2,\mathbb{R})}$ &  $J_3^{\mathbb{O}}$\\ \hline
    8        &  $n+2$  & $\frac{SL(2,\mathbb{R})}{ U(1)} \times \frac{SO(n,2)}{SO(n) \times SO(2)}$ & $\frac{SO(n+2,4)}{SO(n,2) \times SO(2,2)}$ &  $\mathbb{R} \oplus \Gamma_{(1,n-1)} $
\\ \hline \hline
   16        &  $n+2$  & $\frac{SL(2,\mathbb{R})}{U(1)} \times \frac{SO(n-4,6)}{SO(n-4) \times SO(6)} $ & $\frac{SO(n-2,8)}{ SO(n-4,2) \times SO(2,6)}$ &  $\mathbb{R} \oplus \Gamma_{(5,n-5)} $
\\ \hline \hline
24 & 16 & $\frac{SO^*(12)}{ SU(6)\times U(1)} $ & $\frac{E_{7(-5)}}{SO^*(12) \times SL(2,\mathbb{R}}$ &  $J_3^{\mathbb{H}}$ \\ \hline
   32        & 28 & $\frac{E_{7(7)}}{ SU(8)} $ & $\frac{E_{8(8)}}{ SO^*(16)}$ &  $J_3^{\mathbb{O}_s}$ \\ \hline
\end{tabular}
\caption{\label{paraspaces} Above we give the number of supercharges $n_Q$, $4D$ vector fields $n_V$ ,
scalar manifolds of supergravity theories  before and after reduction
along a timelike Killing vector from $D=4$ to $D=3$, and associated Jordan algebras $J$. Isometry groups of $4D$ and $3D$ supergravity theories are given by the conformal, $Conf(J)$, and quasiconformal groups , $QConf(J)$,  of $J$, respectively. }
\end{table}

For dimensionally reducing  the~$D=4$ theory
along a timelike direction one makes the standard  Kaluza-Klein-type
ansatz~\cite{Breitenlohner:1987dg}
\begin{equation} \label{4dmetric}
ds_4^2 = - e^{2U} (dt + \omega)^2 + e^{-2U} ds_3^2\ .
\end{equation}
which results in  $3D$ Euclidean gravity  coupled to scalars, vectors and fermions. Vector fields in three dimensions can be dualized to
scalars and  the bosonic sector is described simply by the three-dimensional
metric~$ds_3^2$ and scalar fields~$\phi^a$.
The three dimensional scalar fields ~$\phi^a$ consist of the scalars coming from the
$4D $ theory, plus electric and magnetic potentials from the reduction of
 $4D$ vector fields ~$A^A_t$ and their duals, plus the scale factor~$U$ and
the twist potential dual to the shift~$\omega$ defined in equation \ref{4dmetric}.
 The resulting  manifold
$\mathcal{M}_3^*$ of scalar fields  can be thought of as  analytic continuation of the
Riemannian manifold~$M_3$ obtained from reduction on a space-like circle.

For spherically symmetric configurations,
the metric on three-dimensional slices can  be written as
\begin{equation}
ds^2_3 = N^2(\rho)\, d\rho^2 + r^2(\rho)\left[ d\theta^2 + \sin^2 \theta\, d \phi^2 \right]\,.
\end{equation}
where $\rho$ is the radial coordinate that  plays the role of time in  radial quantization.
 The bosonic part of the action then becomes
\begin{equation}
\label{sbos}
S = \int d\rho \left[ \frac{N}{2} + \frac{1}{2N}
\left( \dot{r}^2 - r^2 G_{ab} \dot{\phi}^a \dot{\phi}^b \right) \right],
\end{equation}
where the dot denotes  derivative  with respect to $\rho$ and~$G_{ab}$ is the metric on~$\mathcal{M}_3^*$.  Thus   four dimensional
equations of motion are equivalent to geodesic motion of a fiducial
particle on a real cone $\mathbb{R}\times\mathcal{M}_3^*$  over~$\mathcal{M}_3^*$. The equation of motion for the
lapse function ~$N$, which is an auxiliary field,  imposes the Hamiltonian constraint
\begin{equation}
H = p_r^2 - \frac{1}{r^2} G^{ab} p_a p_b - 1 \equiv 0\, ,
\end{equation}
where~$p_r$ and~$p_a$ are the canonical conjugates to~$r$ and
$\phi^a$, respectively.  This constraint fixes the mass of the fiducial particle on the cone
to be ~1. Note that for BPS black holes, one may choose ~$N=1, \rho=r, p_r=1$. With this choice
the problem reduces to {\it light-like} geodesic motion on~$\mathcal{M}_3^*$,
with affine parameter~$\tau=1/r$.
The magnetic and electric charges of the black hole
are simply Noether charges~$P^A$,~$Q_A$ associated with the generators of
$4D $ gauge transformations in the isometry group~$G_3$ acting on
$\mathcal{M}_3^*$.  These charges generate an
Heisenberg subalgebra under Poisson brackets
\begin{equation}
\label{heis}
[P^A, Q_B]_{\rm PB} = 2 \delta^A_B\  K\, ,
\end{equation}
where the "central charge" ~$K$ is the NUT charge of the black hole~\cite{Pioline:2005vi,Gunaydin:2005mx,Gunaydin:2007bg,Gunaydin:2006bz}.\footnote{ The solutions  with~$K\neq 0$ have closed timelike curves
when lifted back to four
dimensions, as a consequence of  the off-diagonal term~$\omega=K\cos\theta d\phi$
in the metric~\eqref{4dmetric}. Therefore  real
 four-dimensional black holes require taking the
"central charge" $K\to 0$ limit.}
The conserved charge of
the isometry that corresponds to  rescalings of the time-time component ~$g_{tt}$ of the metric is the ADM mass $M$ that satisfies
\begin{equation}
[M,P^A]_{\rm PB}=P^A,\ [M,Q_A]_{\rm PB}=Q_A,\ [M,K]_{\rm PB}=2K\, .
\end{equation}

For supergravity theories whose scalar manifolds ~$\mathcal{M}_3^*=G_3/H_3$ are homogeneous or symmetric spaces  there exist
additional conserved charges associated with the additional isometries.
For  $N=2$ MESGTs defined by Euclidean Jordan algebras $J$ of degree three the full isometry group of $\mathcal{M}_3^*$ is the quasiconformal group $QConf(J)$ that has a five grading with respect to the generator $M$
\begin{equation}
\mathfrak{qconf}(J) =  \Tilde{K} \oplus ( \Tilde{P}^A, \Tilde{Q}_A ) \oplus \left( \mathfrak{conf}(J) + M \right) \oplus ( P^A,Q_A) \oplus K
\end{equation}
where $\mathfrak{conf}(J)$ is the Lie algebra of the $4D$ U-duality group $Conf(J)$, which commutes with ADM mass generator $M$.

For spherically
symmetric stationary solutions,
the supersymmetry variation of the fermionic fields $\lambda^\alpha$  are of the general form
\cite{Bagger:1983tt}
\begin{equation}
\delta\lambda^\alpha = V^\alpha_i \epsilon^i,
\end{equation}
where~$\epsilon^i$ is the supersymmetry parameter and~$V^\alpha_i$
is a matrix linear in the velocities~$\dot\phi^a \equiv \frac{\partial \phi}{\partial \tau} $ on~$\mathcal{M}_3^*$.
For general~$\N=2$ MESGTs reduced to $d=3$ , the
indices~$i=1,2$ and~$\alpha=1,\dots, 2n_V+2$ transform as fundamental
representations
of the restricted
holonomy group $Sp(2,\mathbb{R}) \times Sp(2n_V+2,\mathbb{R})$ of para-quaternionic geometry.
For  supersymmetric backgrounds this variation vanishes
for some non-zero~$\epsilon^i$.
One can show that this
is equivalent  to the system of equations~\cite{Pioline:2005vi,Gunaydin:2005mx,Gunaydin:2007bg,Gunaydin:2006bz}:
\begin{eqnarray}
\label{genatt1}
\frac{dz^I}{d\tau} &=& - e^{U+i\alpha} g^{I\bar J} \partial_{\bar J} |Z| \\
\label{genatt2}
\frac{dU}{d\tau} + \frac{i}{2} K &=& - 2 e^{U+i\alpha} |Z|\, ,
\end{eqnarray}
where
\begin{equation}
Z(P,Q,K) = e^{\mathcal{K}/2} \left[ (Q_A - 2 K \tilde{\zeta}_A) X^A - (P^A + 2K\zeta^A) F_A \right]
\end{equation}
is the central charge function\footnote{
The phase~$\alpha$ is to be chosen such that
$dU/d\tau$ is real.}.

For vanishing NUT charge $K$ , the above equations take the form of
the standard  attractor flow equations describing the radial evolution of the scalars towards the black hole horizon ~\cite{Ferrara:1995ih,
Ferrara:1996um,Ferrara:1997tw,Moore:1998pn,Denef:2000nb}
\begin{eqnarray}
\frac{dU}{d\tau} &=& - 2 e^{U} |Z|  \\
\frac{dz^I}{d\tau} &=& - ~e^U~ g^{I\bar{J}}
\partial_{\bar{J}} |Z|\
\end{eqnarray}
 with the central charge function \[ Z(P,Q, K=0) = e^{\mathcal{K}/2} \left[ Q_A  X^A - P^A  F_A \right] \]
 The equivalence of attractor flow of $N=2$ supergravity in $d=4$ and supersymmetric geodesic motion on $M_3^*$ was  pointed out  in~\cite{Gutperle:2000ve}.

 The scalar fields $\tilde{\zeta}_A,\zeta^A$  conjugate to the charges $P^A$ and $Q_A$ evolve according to
\begin{eqnarray}
\frac{d\zeta^A}{d\tau} &=&
-\frac12 e^{2U} [(Im {\cal N})^{-1}]^{AB} \nonumber\\
&&\left[ Q_A - 2 K \tilde{\zeta}_A - [Re {\cal N}]_{BC} ( P^C + 2 K \zeta^C)
\right]\\
\frac{d\tilde{\zeta}_A}{d\tau} &=&
-\frac12 e^{2U} [Im {\cal N}]_{AB} ( P^B + 2 K \zeta^A)
-[Re {\cal N}]_{AB}
\frac{d\zeta^J}{d\tau} \nonumber
\end{eqnarray}
where ${\cal N}_{AB}$ is the period matrix of special geometry~\cite{Ceresole:1995ca}.

For ~$\N=2$ MESGTs defined by Euclidean Jordan algebras $J$ of degree three the holonomy group of $\mathcal{M}_3^*$ is
~$Conf(J)  \times Sp(2,\mathbb{R}) \subset Sp(2n_V+4,\mathbb{R})$.
The full phase space is $8n_V+8 $ dimensional and for BPS black holes supersymmetry leads to $ 2n_V+1$ first class constraints which reduce the dimension of the phase space to $4n_V+6= (8n_V+8) - 2(2n_V+1) $. This reduced phase of BPS black holes can be identified with the twistor space of $\mathcal{M}_3$ of complex dimension $(2n_V+3)$ \cite{Gunaydin:2005mx,Gunaydin:2007bg,Gunaydin:2007qq,Neitzke:2007ke}.

The twistor space of the scalar manifold
\begin{equation*}
 \mathcal{M}_3=  \frac{QConf(J)}{\widetilde{Conf}(J)\times SU(2)}
\end{equation*}
of dimensionally  reduced $N=2$ MESGT defined by a Jordan algebra $J$ is
\begin{equation}
\mathcal{Z}_3 = \frac{QConf(J)}{\widetilde{Conf}(J)\times SU(2)} \times \frac{SU(2)}{U(1)} = \frac{{QConf}(J)}{\widetilde{Conf}(J) \times U(1)} \label{twistorspace}
\end{equation}

The quasiconformal group action of a group $G$  
extends to its complexification \cite{Gunaydin:2000xr}. Consequently, quasiconformal
 actions of three dimensional U-duality groups $QConf(J)$ on the space with real coordinates
 $\mathcal{X} = (X,x)$
 extend
naturally to the complex coordinates $\mathcal{Z}= (Z,z)$ of corresponding twistor spaces $\mathcal{Z}_3$ \cite{Gunaydin:2007qq}. The K\"ahler potential of the K\"ahler-Einstein metric of the twistor space is given precisely by the "quartic light-cone" of quasiconformal geometry in these coordinates
\begin{equation}
\mathcal{K}(\mathbf{\mathcal{Z}},\mathbf{\mathcal{\bar{Z}}} )= \ln
 d\left(\mathbf{\mathcal{Z}}, \mathbf{\mathcal{\bar{Z}}}\right) = \ln
 \left[ \mathcal{Q}_4 (\mathbf{Z} - \mathbf{\bar{Z}} ) + \left(
 \mathbf{z} - \mathbf{\bar{z}} + \langle \mathbf{Z} , \mathbf{\bar{Z}}
 \rangle \right)^2\right] \label{Kahlerpotential}
\end{equation}
The K\"ahler potential  is manifestly invariant under the Heisenberg
symmetry group corresponding to ``symplectic translations'' generated
by $Q_A$, $P^A$  and $K$.  Under the global action of ``symplectic special conformal
generators'' $\Tilde{Q}_A$, $\Tilde{P}^A$ and $\Tilde{K}$ the quartic light-cone transforms as\cite{Gunaydin:2000xr,Gunaydin:2007qq}
\begin{equation}
 d\left(\mathbf{\mathcal{Z}}, \mathbf{\mathcal{\bar{Z}}}\right)
  \Longrightarrow f(\mathbf{Z},\mathbf{z}) \bar{f}(
  \mathbf{\bar{Z}},\mathbf{\bar{z}} ) d\left(\mathbf{\mathcal{Z}},
  \mathbf{\mathcal{\bar{Z}}}\right)
\end{equation}
which correspond to   K\"ahler transformations of the K\"ahler
potential \ref{Kahlerpotential} of the twistor space and
hence leaves the K\"ahler metric invariant.  These results were first established for
quaternionic symmetric spaces \cite{Gunaydin:2007qq} and  there exist analogous  K\"ahler potentials for more general quaternionic manifolds  invariant only under the Heisenberg
symmetries generated by $Q_A$ , $P^A$ and $K$ that are in the C-map\cite{Neitzke:2007ke}.  As will be discussed in the next section,  the correspondence established
between harmonic superspace formulation of $4D$, $N=2$ sigma models coupled
to $N=2$ supergravity and quasiconformal realizations of their isometry
groups \cite{Gunaydin:2007vc} implies that  K\"ahler potentials of
quartic light-cone type must  exist for all quaternionic target manifolds  and not only those that are in the C-map.

The quantization of the motion of fiducial particle on $\mathcal{M}_3^*$ leads to quantum
mechanical wave functions that provide the basis of a unitary
representation of the isometry group $G_3$ of $\mathcal{M}_3^*$ .  BPS black holes correspond to a special
class of geodesics which lift holomorphically to the twistor space
$\mathcal{Z}_3$ of $\mathcal{M}_3^*$.  Spherically symmetric stationary
BPS black holes of $N=2$ MESGT's are described by holomorphic curves
in $\mathcal{Z}_3$
\cite{Gunaydin:2005mx,Gunaydin:2007bg,Gunaydin:2007qq,Neitzke:2007ke}.
Therefore for theories defined by Jordan algebras $J$ of degree three, the relevant unitary representations of the isometry groups $QConf(J)$  for BPS
black holes are those induced by their holomorphic actions on the corresponding twistor spaces $\mathcal{Z}_3$, which
belong in general to quaternionic discrete series representations \cite{Gunaydin:2007qq}.
For rank two quaternionic groups $SU(2,1)$ and $G_{2(2)}$ unitary representations induced by the geometric quasiconformal actions were studied in great detail in \cite{Gunaydin:2007qq}.

\section{Harmonic Superspace, Minimal Unitary Representations  and  Quasiconformal Group Actions}
\renewcommand{\theequation}{\arabic{section}.\arabic{subsection}.\arabic{equation}}
\setcounter{equation}{0}

In this section we shall review  the   connection
between the harmonic superspace (HSS) formulation of  $4D$, $N=2$ supersymmetric quaternionic K\"ahler sigma models that couple to $N=2$ supergravity and the minimal unitary representations of their
isometry groups \cite{Gunaydin:2007vc}.  We shall then discuss the relevance of these results to the proposal that
      quasiconformal extensions  of  U-duality groups of four dimensional   $N=2$ MESGTs  must act  as spectrum generating
 symmetry groups \cite{Gunaydin:2000xr,Gunaydin:2005gd,Gunaydin:2005mx,Gunaydin:2007bg,Gunaydin:2007qq}, which extends the proposal that the conformal extensions of  U-duality groups of
 $N=2$, $d=5$ MESGTs act as spectrum generating symmetry groups \cite{Ferrara:1997uz,Gunaydin:2000xr,Gunaydin:2005gd}.

\subsection{$4D$,  $N=2$ $\sigma$-models coupled to Supergravity  in Harmonic Superspace}

The target spaces of $N=2$ supersymmetric $\sigma$-models coupled to
$N=2$ supergravity in four dimensions are quaternionic K\"ahler manifolds \cite{Bagger:1983tt}. They can be formulated in a manifestly supersymmetric form in harmonic superspace
\cite{Galperin:1984av,Bagger:1987rc,Galperin:1992pj,Galperin:1992pw} which we shall review briefly following \cite{Galperin:1992pw}.  In harmonic superspace approach  the  metric on a quaternionic target space of $N=2$ sigma model is
given  by a  quaternionic potential ${\cal L}^{(+4)}$, which  is the analog of   K\"ahler potentials of complex K\"ahler manifolds.

The $N=2$ harmonic superspace action for the general $4n$-dimensional quaternionic $\sigma$-model
has the simple form \cite{Galperin:1992pw}\footnote{ The conventions for indices in this section are independent of the conventions of previous sections and follow closely the conventions used in \cite{Gunaydin:2007vc}. The number of plus (+)  or  minus (-)  signs in a superscript or subscript  denote the $U_A(1)$ charges.  If the $U_A(1)$  charge $ n > 2$ ,  it is indicated as $(+n)$. } 
\eq
 S=\int d\zeta^{(-4)}du \{ Q_\alpha^+D^{++}Q^{+\alpha} -q^+_i D^{++} q^{+i} + \lpr
(Q^+,q^+,u^-)\}.\label{action}\en where the integration is over the {\it analytic} superspace coordinates
$\zeta, u^{\pm}_i$. The  hypermultiplet superfields $Q^+_\alpha(\zeta, u), \; \alpha =1,...,2n$ and the supergravity
hypermultiplet compensators $q^+_i(\zeta, u), \;(i=1,2)$ are analytic $N=2$ superfields. The   $
u^\pm_i,(i=1,2)$ are the $S^2= \frac{SU_A(2)}{U_A(1)}$ isospinor harmonics that satisfy \[
u^{+i}u^-_i=1
\] and $D^{++}$ is a supercovariant derivative with respect to harmonics with the property
\[ D^{++}u^-_i=u^+_i \]

The analytic subspace of the full $N=2$ harmonic superspace involves only half the Grassmann
variables with coordinates $\zeta^M$ and  $u^{\pm}_i $

\eq \zeta^M := \{ x^\mu_A , \;  \theta^{a +}
, \; \bar{\theta}^{\dot{a}+} \} \en
 where \[ x^\mu_A := x^\mu -2i \theta^{(i} \sigma^\mu
\bar{\theta}^{j)} u_i^+ u_j^- \]
\[ \theta^{a +} := \theta^{a i} u_i^+ \]
\[ \bar{\theta}^{\dot{a}+} :=\bar{\theta}^{\dot{a} i} u_i^+ \]
\[ \theta^{(i} \sigma^\mu \bar{\theta}^{j)} u_i^+ u_j^- := \theta^{(ai} (\sigma^\mu)_{a\dot{a}}
\bar{\theta}^{\dot{a}j)} u_i^+ u_j^- \]
\[ \mu=0,1,2,3 ; \; a=1,2 ; \; \dot{a}=1,2 \]

Furthermore the analytic subspace does not involve $U(1)$ charge $-1$ projections of the Grassmann variables
and still closes  under $N=2$ supersymmetry transformations.  It satisfies a  "reality condition" with respect to
the conjugation $\,\widetilde{}$
\begin{eqnarray}
\widetilde{x}^\mu=x^\mu  \nonumber \\
\widetilde{\theta}^+ = \bar{\theta}^+ \\
\widetilde{\bar{\theta}}^+ = - \theta^+ \nonumber \\
\widetilde{u}^{i\pm} = u_i^{\pm}  \nonumber \\
\widetilde{u}_i^{\pm} - - u^{\i\pm}  \nonumber
\end{eqnarray}
which is a product of  complex conjugation and anti-podal map on the sphere $S^2$ \cite{Galperin:2001uw}.

The quaternionic potential $\lpr$ is a homogeneous function in $Q^+_\alpha$  and $ q^+_i$ of degree two and
has  $U(1)$-charge $+4$.  It does not depend on $u^+$ and  is, otherwise,  an arbitrary "real function"
with respect to the involution $\widetilde{}$. We shall suppress the dependence of all the fields on the harmonic superspace
coordinates $\zeta^M$ and $u^{\pm}_i$ in what follows.

The action \ref{action} is of the form of Hamiltonian mechanics with the harmonic derivative $D^{++}$ playing the
role of  time derivative  \cite{Galperin:1991rk,Galperin:1992pw,Galperin:2001uw} and with  the superfields $Q^+$ and $ q^+$ corresponding  to phase space coordinates under the Poisson brackets
\eq \{f,g\}^{--} = {1\over 2}\Omega^{\alpha\beta}{\pa f\over \pa Q^{+\alpha}} {\pa
g\over \pa Q^{+\beta}}- {1\over 2}\epsilon^{ij}{\pa f\over\pa q^{+i}} {\pa g\over\pa q^{+j}},
\label{pb}\en
$\Omega^{\alpha\beta}$ and $\epsilon^{ij}$ are the invariant antisymmetric
tensors of $Sp(2n)$ and $Sp(2)$ , respectively, which  are used to raise and lower indices
\eq Q^{+\alpha}= \Omega^{\alpha\beta}Q^+_\beta \en \[ q^{+i}=\epsilon^{ij}q^+_j \] and
satisfy\footnote{ Note that the conventions of \cite{Galperin:1992pw} for the symplectic metric ,which we follow in this section, differ
from those of \cite{Gunaydin:2006vz}.}

\eq \Omega^{\alpha\beta}\Omega_{\beta\gamma} = \delta^{\alpha}_{\gamma} \en \eq \epsilon^{ij}
\epsilon_{jk}= \delta^i_k \en
The quaternionic potential
$\lpr$ plays the role of the  Hamiltonian and  we shall refer to it as such following \cite{Galperin:1992pw}.

Isometries of the $\sigma$-model \ref{action} are generated by Killing potentials $K^{++}_A(Q^+, q^+,
u^-)$ that obey the conservation law \eq \pa^{++}K^{++}_A +\{K^{++}_A, \lpr\}^{--}=0\
\label{cons}\en
 where $\partial_{++}$ is defined as
\[\partial_{++} = u^{+i} \frac{\partial}{\partial u^{-i}} \]
They generate  the Lie algebra of the isometry group \eq \{K^{++}_A,
K^{++}_B\}^{--}=f_{AB}^{\ \ \ C}K^{++}_C\ .\label{Lie} \en under  Poisson brackets \ref{pb}.

The "Hamiltonians" $\lpr$ of all $N=2$ $\sigma$-models  coupled to N =2 supergravity with irreducible symmetric target
manifolds were given in \cite{Galperin:1992pw}. The quaternionic symmetric spaces , sometimes known as Wolf
spaces  \cite{MR0185554}, that arise in $N=2$  supergravity  are of the non-compact type. For each
simple Lie group  there is a unique non-compact quaternionic symmetric space. Below we give  a complete list of
these spaces:
\eq\begin{array}{c} \fracds{SU(n,2)}{U(n) \times
Sp(2)} \\ \fracds{SO(n,4)}{SO(n)\times SU(2) \times Sp(2)}  \\ \fracds{USp(2n,2)}{Sp(2n)\times
Sp(2)} \\
\fracds{G_{2(+2)}}{SU(2)\times Sp(2)}\\
\fracds{F_{4(+4)}}{Sp(6)\times Sp(2)} \\ \fracds{E_{6(+2)}}{SU(6)\times Sp(2)} \\
\fracds{E_{7(-5)}}{SO(12)\times Sp(2)}  \\ \fracds{E_{8(-24)}}{E_7 \times Sp(2)}.
\end{array}
\label{one} \en

Given  a target space $\frac{G}{H\times Sp(2)}$ of $N=2$ $\sigma$ model
coordinatized by $Q^+_\alpha$ and  $q^+_i$ , every generator $\G_A$ of $G$ maps to a function $K^{++}_A(Q^+, q^+,
u^-)$ such that the action of $K^{++}_A$ is given via the Poisson brackets \ref{pb}. Furthermore,  it can be shown that
  the Hamiltonian $\lpr$ depends only on $Q^+_\alpha$ and the combination
$(q^+u^- )\equiv q^{+i}u_i^-$ \cite{Galperin:1992pw} 
 \eq \lpr =\lpr (Q^+,(q^+u^-)) .\label{qu} \en
and can be written as
\eq \lpr
={P^{(+4)}(Q^+)\over (q^+u^-)^2} \label{formofl}\en
The fourth order polynomial $P^{(+4)}$ is given by
\eq P^{(+4)}(Q^+)=\frac{1}{12} \; S_{\alpha\beta\gamma\delta}\;
Q^{+\alpha}Q^{+\beta}Q^{+\gamma}Q^{+\delta}\label{P^4} \en where $S_{\alpha\beta\gamma\delta}$  is
a completely symmetric invariant tensor of $H$. In terms of  matrices $t^a_{\alpha\beta} =t^a_{\beta\alpha}, \; a=
1,2,.., dim(H)$
given by the action of the generators $K^{++}_a$ of $H$ on the coset space generators $ K^{++}_{ i \alpha}$
\eq
\{K^{++}_a , K^{++}_{i \alpha} \} = t_{a \alpha}^\beta  K^{++}_{ i \beta}
\en
the invariant tensor reads as \cite{Galperin:1992pw} \eq   S_{\alpha\beta\gamma\delta}= h_{ab} t^a_{\ \alpha\beta}t^b_{\ \gamma \delta}
+\Omega_{\alpha\gamma}\Omega_{\beta\delta}+\Omega_{\alpha\delta}\Omega_{\beta\gamma} \ .\label{z12}\en where $h_{ab}$ is the Killing metric of
$H$.

The Killing potentials that generate the isometry group $G$ are given by \cite{Galperin:1992pw}
\eq
\mathbf{Sp(2)}:\;\;\;\; K^{++}_{ij}=2(q^+_iq^+_j-u^-_iu^-_j\lpr), \label{su2curr} \en
\eq \mathbf{H}:\;\;\;\;
K^{++}_a=t_{a\alpha\beta}Q^{+\alpha}Q^{+\beta},\label{Hcurr} \en
\eq \mathbf{G/H\times
Sp(2)}:\;\;\;\;K^{++}_{i\alpha}=2q^+_iQ^+_\alpha - u^-_i(q^+u^-) \pa^-_\alpha\lpr\
,\label{cosetcurr} \en
where
\[ \pa^-_\alpha :=\frac{\partial}{\partial Q^{+\alpha}} \]
\[ t_{a\a\b}=\Omega_{\beta\gamma}t_{a\alpha}^{\ \ \ \gamma} \]
The Killing potentials $K^{++}_{ij}$ that generate $Sp(2)$ are conserved for any
arbitrary polynomial $P^{(+4)}(Q^+)$.
 Since $t^a$ are also the representation matrices of the generators of $H$ acting on $Q^{+\alpha}$ one finds
that
the fourth order polynomial  $P^{(+4)}$  is proportional to the quadratic "Casimir function"
$h^{ab}K_a^{++}K_b^{++}$ of $H$. Furthermore , $P^{(+4)}$ can also be expressed in terms of
Killing potentials of the coset space generators, or in terms of the $Sp(2)$ Killing potentials  as follows \cite{Galperin:1992pw} : \eq
P^{(+4)}=-\frac{1}{16}\e^{ij}\Omega^{\alpha\beta}K^{++}_{i\a} K^{++}_{j\b}=-\frac{1}{8}K^{++ij}K^{++}_{ij}\ .\en

\subsection{ Minimal Unitary Representations
of Non-compact Groups from their Quasiconformal Realizations}

Before discussing  its relationship to the HSS formulation of $N=2$ sigma models given in the previous section, we shall review the unified construction of  minimal unitary representations
of noncompact groups obtained by quantization of their
geometric realizations as quasiconformal groups following \cite{Gunaydin:2006vz} which generalizes the
results of \cite{Gunaydin:2001bt,Gunaydin:2004md,Gunaydin:2005zz}.

Consider the 5-graded decomposition of  the Lie algebra
$\mathfrak{g}$ of a noncompact group $G$ of quaternionic type and label the generators such that 

\begin{equation}
\mathfrak{g}=  \mathfrak{g}^{-2}  \oplus  \mathfrak{g}^{-1}  \oplus \left(  \mathfrak{h}  \oplus \Delta
\right)
 \oplus  \mathfrak{g}^{+1}  \oplus  \mathfrak{g}^{+2} \nonumber
\end{equation}
\eq \mathfrak{g} = E \oplus E^\alpha \oplus ( J^a + \Delta ) \oplus F^\alpha \oplus F \en
where $\Delta$ is the generator that determines the 5-grading.
Generators $J^a$  of  $\mathfrak{h}$ satisfy
\begin{subequations}\label{eq:alg}
\begin{equation}
  \left[ J^a \,, J^b \right] = {f^{ab}}_c J^c
\end{equation}
where $a,b,...=1,...D=dim(H)$.  We shall denote  the symplectic
representation by which $\mathfrak{h}$ acts on the subspaces $\mathfrak{g}^{\pm
1}$ as $\rho$
\begin{equation}
  \left[ J^a \,, E^{\alpha} \right] = {\left(\lambda^{a}\right)^\alpha}_\beta E^\beta ,
\qquad
  \left[ J^a \,, F^{\alpha} \right] =  {\left(\lambda^{a}\right)^\alpha}_\beta F^\beta
\end{equation}
where $E^\alpha$, $ \alpha, \beta, ..= 1,..,N= \dim (\rho)$ are
generators that span the subspace $\mathfrak{g}^{-1}$
\begin{equation}
  \left[ E^\alpha \,, E^\beta  \right] = 2 \Omega^{\alpha\beta} E
\end{equation}
and $F^\alpha$ are generators that span $\mathfrak{g}^{+1}$
\begin{equation}
  \left[ F^\alpha \,, F^\beta  \right] = 2 \Omega^{\alpha\beta} F
\end{equation}
$\Omega^{\alpha\beta}$ is the symplectic invariant ``metric'' of
the representation $\rho$.
The remaining nonvanishing commutation relations of $\mathfrak{g}$ are given by
\begin{equation}
\begin{aligned}
 F^\alpha &= \left[ E^\alpha \,, F \right] \cr
 E^\alpha &= \left[ E \,, F^\alpha \right] \cr
 \left[E^{\alpha} , F^{\beta}\right] &= - \Omega^{\alpha\beta} \Delta + \epsilon \lambda_a^{\alpha\beta}
J^a
\end{aligned}
\qquad
\quad
\begin{aligned}
\left[\Delta, E^{\alpha} \right] &= - E^{\alpha} \cr
\left[\Delta, F^{\alpha} \right] &= F^{\alpha} \cr
\left[\Delta, E\right] &= -2 E \cr
\left[\Delta, F \right] &= 2 F
\end{aligned}
\end{equation}
\end{subequations}
where
$\epsilon$ is a  constant parameter whose value depends on the Lie algebra $\mathfrak{g}$.
Note that the positive  (negative)   grade generators form a
Heisenberg subalgebra since
\begin{equation}
 \left[E^{\alpha}, E \right] = 0
\end{equation}
with the grade +2 (-2) generator $F$ ($E$)   acting as its central charge.

In the minimal unitary realization of noncompact groups \cite{Gunaydin:2006vz}, negative grade
  generators are expressed as bilinears of  symplectic bosonic oscillators
$\xi^\alpha$ satisfying the canonical commutation relations\footnote{ In this section  we follow the
conventions of \cite{Gunaydin:2006vz}.  The indices $\alpha, \beta,..$ are raised and lowered with the
antisymmetric symplectic metric $\Omega^{\alpha\beta}=-\Omega^{\beta\alpha}$ that satisfies
$\Omega^{\alpha\beta}\Omega_{\gamma\beta}=\delta^{\alpha}_{\beta} $ and $V^\alpha =
\Omega^{\alpha\beta} V_\beta $, and $V_\alpha =V^\beta \Omega_{\beta\alpha} $.}
\begin{equation}
  \left[ \xi^\alpha \,, \xi^\beta \right] = \Omega^{\alpha\beta}
\end{equation}
and an extra coordinate $y$ \footnote{ In the corresponding  geometric quasiconformal realization of the group $G$ over an $(N+1) $ dimensional space  this coordinate corresponds to the singlet  of $H$ }.

\begin{equation}
 E = \frac{1}{2} y^2 \qquad E^\alpha = y \,\xi^\alpha \qquad
J^a =  - \frac{1}{2}  {\lambda^a}_{\alpha\beta} \xi^\alpha \xi^\beta
\end{equation}

The quadratic Casimir operator of the Lie algebra $\mathfrak{h}$ is
\begin{equation}
\mathcal{C}_2\left(\mathfrak{h}\right) = \eta_{ab} J^a J^b
\end{equation}
where $\eta_{ab}$ is the Killing metric of the subgroup  $H$ , which is isomorphic to the
automorphism group of the underlying FTS $\mathcal{F}$. The quadratic Casimir $C_2(\mathfrak{h})$
is equal to the quartic invariant of $H$ in the representation $\rho$ modulo an additive constant
that depends on the normal ordering chosen, namely
\eq I_4(\xi^\alpha) = S_{\alpha\beta\gamma\delta} \xi^\alpha \xi^{\beta} \xi^{\gamma} \xi^{\delta}
=  C_2(\mathfrak{h}) + \mathfrak{c}\en
where $\mathfrak{c}$ is a  constant and \[ S_{\alpha\beta\gamma\delta} := \lambda_{a(\alpha\beta}
\lambda^a_{\gamma\delta)}\]

The grade +2 generator $F$ has the general form
\begin{equation} \label{eq:exprF}
  F = \frac{1}{2} p^2 + \frac{\kappa  \left( \mathcal{C}_2(\mathfrak{h}) + \mathfrak{C} \right)}{y^2}
\end{equation}
where $p$ is the momentum conjugate to the singlet coordinate $y$
\begin{equation}
[y,p]=i
\end{equation}
and $\kappa$ and $\mathfrak{C}$ are some constants depending on the Lie algebra $\mathfrak{g}$.
The grade $+1$ generators are then given  by commutators
\begin{equation}
 F^\alpha = \left[E^{\alpha}, F\right]= i p \, \xi^\alpha + \kappa y^{-1} \left[ \xi^\alpha \,, \mathcal
{C}_2
 \right]
 \end{equation}
 If $H$ is  simple or Abelian  they take the general form \cite{Gunaydin:2006vz}
 \begin{equation}
F^\alpha= ip \, \xi^\alpha -\kappa y^{-1} \left[ 2 \, {\left(\lambda^{a}\right)^{\alpha}}_\beta \xi^\beta
J_a
 + \, C_\rho \,\xi^\alpha \right]
\end{equation}
where $C_\rho$ is the eigenvalue of the second order Casimir of $H$
in the representation $\rho$ . Furthermore,  one finds
\begin{equation}
 \left[ E^\alpha \,, F^\beta \right] = - \Delta \Omega^{\alpha\beta} - 6 \kappa \left(\lambda^a\right)^
{\alpha\beta} J_a
\end{equation}
where $\Delta = - \frac{i}{2} \left( yp+py\right)$\footnote{ In the most general case, where $H$ is not necessarily simple or Abelian , one finds that the
commutator of $ E^\alpha $ and $ F^\beta $ has the same form as above.}.

For simple $H$ the quadratic Casimir operator of the Lie algebra $\mathfrak{g}$ can be calculated simply   \cite{Gunaydin:2006vz}
\begin{equation}
 \mathcal{C}_2 \left( \mathfrak{g} \right) = J^a J_a + \frac{2 \, C_\rho}{N+1} \left( \frac{1}{2}\,
\Delta^2 + E F + F E \right) - \frac{ C_\rho}{N+1}
 \Omega_{\alpha\beta} \left( E^\alpha F^\beta + F^\beta E^\alpha \right)
\end{equation}
The quadratic Casimir of $sl(2)$ generated by $E,F$ and $\Delta$ can be expressed in terms of
the quadratic Casimir $J^aJ_a$  of $H$ :
\begin{equation}
  \frac{1}{2}\, \Delta^2 + E F + F E = \kappa\left( J^a J_a + \mathfrak{C} \right) - \frac{3}{8}
  \en
as well as  the contribution of the coset
generators $F^\alpha$ and $E^\beta$ to  $\mathcal{C}_2 \left( \mathfrak{g} \right)$ to the Casimir of $\mathfrak{g}$
\eq
  \Omega_{\alpha\beta}\left( E^\alpha F^\beta + F^\beta E^\alpha \right) = 8 \, \kappa J^a J_a + \frac{N}
{2} + \kappa C_\rho N
\end{equation}
Thus the quadratic Casimir of $\mathfrak{g}$ reduces to a c-number\cite{Gunaydin:2006vz}
\begin{equation}
\mathcal{C}_2 \left( \mathfrak{g} \right) =  \mathfrak{C} \left( \frac{8 \kappa C_\rho}{N+1} - 1 \right) -
\frac{3}{4} \, \frac{C_\rho}{N+1} - \frac{N}{2} \, \frac{C_\rho}{N+1} - \frac{\kappa C_\rho^2 N}{N+1}
\end{equation}
as required by irreducibility of the minimal representation and is a general  feature for  minimal unitary realizations of
all simple groups $G$ obtained from their quasiconformal realizations\cite{Gunaydin:2001bt,Gunaydin:2004md,Gunaydin:2005zz,Gunaydin:2006vz}.

\subsection{Harmonic Superspace   Formulation of $N=2$ Sigma Models and Minimal Unitary Representations of  Their Isometry Groups}

Let us now  show that there is  a precise mapping between the Killing potentials of the isometry group $G$  of the sigma
model in harmonic superspace and the generators of its minimal unitary realization \cite{Gunaydin:2007vc}. This is best achieved
by rewriting the Killing potentials in terms of  $SU(2)_A$ invariant canonical variables, which are defined as follows
\eq
\sqrt{2} q^{+i} u_i^- := w  \en
\eq
\sqrt{2} q^{+i} u_i^+ := p^{++}
\en
The poisson brackets of $q^{+i}$
\eq \{q^{+i},q^{+j}\} = -\frac{1}{2} \epsilon^{ij} \en
imply that
\eq \{w,p^{++}\}= -1 \en
Under the conjugation $\widetilde{}$ we have \[ \widetilde{\widetilde{q^{+i}}} = - q^{+i} \]
\[ \widetilde{\widetilde{u^{\pm}_i }} = - u_i^{\pm} \]
which imply \eq \widetilde{\widetilde{ w}} = w \en
\eq \widetilde{\widetilde{p^{++}}} = p^{++} \en
The Hamiltonian  can then be written as
 \eq \lpr ={2P^{(+4)}(Q^+)\over w^2}
\label{formofl}
\en

The $SU_A(2)$ invariant Killing potentials that generate the isometry group $G$ are then

\eq
\begin{array}{lll}
\mathbf{Sp(2)}: & M^{(+4)}&:= M^{++++}  = K^{++}_{ij}u^{+i}u^{+j} = (p^{++})^2 - \frac{2P^{(+4)}(Q^+)}{w^2}  \\
& M^{++} &= K^{++}_{ij} ( u^{+i}u^{-j} + u^{+j} u^{-i} )= w p^{++} +p^{++} w \\
& M^0 &= K^{++}_{ij} u^{-i}u^{-j} = w^2 \\
&&\\
\mathbf{H}:& K^{++}_a & =t_{a\alpha\beta}Q^{+\alpha}Q^{+\beta}  \\
&& \\
\mathbf{G/H\times Sp(2)}: & T^{(+3)} &:=T^{+++}_{\alpha}  = K^{++}_{i\alpha} u^{+i}= -\sqrt{2} \{ p^{++} Q^+_\alpha -
\frac{\pa^-_\alpha  P^{(+4)}(Q^+) }{w}  \}  \\
&  T^+_{\alpha} & = K^{++}_{i\alpha} u^{-i}= -\sqrt{2} w Q_{\alpha}^+ 
\end{array}
\en

Comparing the above Killing potentials of the isometry group $G$ with the generators of the minimal
unitary realization of $G$ given above we have   one-to-one correspondence between the elements of
harmonic superspace (HSS) and those of  minimal unitary realizations (MINREP) given in Table \ref{map}

\begin{table}
 \begin{tabular}{|| c | c ||}
\hline
\hline
 $\sigma$-model with Isometry  Group $G$ in HSS    &  Minimal Unitary Representation of $G$ \\ \hline
  $w$ & $y$ \\
& \\ \hline
 $ p^{++} $  & $ p $ \\
& \\ \hline
   $\{ \;,\;\}$ & $i[\;,\;]$ \\
& \\ \hline
   $Q^{+\alpha}$ & $\xi^{\alpha}$ \\
& \\ \hline
   $P^{(+4)}(Q^+)$ & $I_4(\xi)$ \\
& \\  \hline
  $ K^{a++}=t^a_{\alpha\beta} Q^{+\alpha} Q^{+\beta}$  & $J^a =\lambda^a_{\alpha\beta} \xi^{\alpha}\xi^
{\beta}$ \\
& \\  \hline
  $T_\alpha^{+++}=K^{++}_{i\alpha} u^{+i}$  & $F^\alpha$ \\
& \\  \hline
 $ T_\alpha^+=K^{++}_{i\alpha} u^{-i}$  & $E^{\alpha}$ \\
& \\  \hline
$M^{++++} $ & $ F $ \\
& \\ \hline
$M^0 $ & $ E $\\
& \\ \hline
$M^{++} $ & $ \Delta $ \\
& \\ \hline
   \hline
 \end{tabular}
\caption{\label{map}  Above we give the correspondence between the quantities of  the harmonic space formulation of  $N=2$ sigma models coupled to supergravity and the operators that enter in the minimal unitary realizations of their isometry groups. }
\end{table}

The Poisson brackets (PB) $ \{ , \}$ in HSS formulation go over to $i$ times the commutator
$ [ , ]$ in the minimal unitary realization and the
harmonic superfields  $w, p^{++}$ corresponding to supergravity hypermultiplet compensators , that are canonically conjugate under PB map to the canonically
conjugate coordinate and momentum operators $y,p$. Similarly, the harmonic superfields $Q^{+\alpha}$ that
form $N/2$ conjugate pairs under Poisson brackets map into the symplectic bosonic oscillators $\xi^\alpha$ on the MINREP side. One finds a normal ordering ambiguity in the quantum versions of the quartic invariants.
The classical expression relating the quartic invariant polynomial $P^{(+4)}$ to the quadratic Casimir
function in HSS formulation  differs from the expression relating the quartic quantum invariant $I_4$ to the quadratic Casimir
 of $H$ by an additive c-number depending on the ordering chosen. The consistent choices for the
 quadratic Casimirs  for all  noncompact groups of quaternionic type were given  in \cite{Gunaydin:2004md,Gunaydin:2005zz,Gunaydin:2006vz}.

Furthermore, the   mapping between HSS formulation of $N=2$  sigma models and minimal unitary realizations  of their isometry groups $G$  extends also to the equations relating the quadratic Casimir of the subgroup  $H$ to the
 quadratic Casimir of $Sp(2)$ subgroup and to the contribution of the coset generators $G/H\times Sp(2)$ to
the quadratic Casimir of $G$ modulo some additive constants due to normal ordering.

On the MINREP side we are working with a realization in terms of  quantum mechanical coordinates and momenta, while
in HSS side the corresponding quantities are classical harmonic analytic superfields. The above correspondence can be extended to the full quantum correspondence on both sides by reducing the $4D$ $N=2$ $\sigma$ model to one dimension
and quantizing it to get a supersymmetric quantum mechanics ( with 8 supercharges). The bosonic  spectrum of the corresponding quantum mechanics must furnish a minimal unitary
representation of the isometry group , which extends to a  fully supersymmetric spectrum.
\subsection{ Implications }

The mapping between the formulation of  $N=2$ , $d=4$ quaternionic K\"ahler $\sigma$ models in HSS and the minimal unitary
realizations of their isometry groups reviewed above is quite remarkable.
It  implies that the {\it fundamental spectra } of the
quantum $N=2$ , quaternionic K\"ahler $\sigma$ models in $d=4$  and their lower dimensional counterparts must fit  into the
minimal unitary representations of their isometry groups.
The fundamental spectra consist of  states created by the action of  harmonic analytic superfields at a
given point in analytic superspace with coordinates $\zeta^M$ on the vacuum of the theory. The above analysis
shows that the states created by  purely bosonic components  of  analytic superfields will
fit into the minimal unitary representation of the corresponding isometry group. Since the analytic
superfields are unconstrained, the bosonic spectrum must extend to full  $N=2$ supersymmetric spectrum ( 8 supercharges) by the
action of the fermionic components of the superfields.

The minimal unitary representations for general noncompact groups are the analogs of the singleton representations of symplectic groups $Sp(2n, \mathbb{R})$ \cite{Gunaydin:2006vz}.  The singleton realizations of $Sp(2n, \mathbb{R})$  correspond to  free field realizations , i.e. their generators can be written as bilinears of bosonic oscillators \footnote{ For symplectic groups the quartic invariant vanishes and hence the minimal unitary realization becomes a free field construction.}.
As a consequence the tensoring procedure becomes simple and straightforward for the symplectic groups \cite{Gunaydin:1981yq}.
However , in general, the minimal unitary realization of a noncompact group is "interacting" and the corresponding
generators are nonlinear in terms of the coordinates and momenta , which  makes the tensoring problem highly nontrivial.  For the quantum $N=2$ quaternionic K\"ahler $\sigma$ models  one then has to
tensor  the fundamental supersymmetric spectra with each other repeatedly to obtain the full 'perturbative" spectra.
The full "nonperturbative" spectra in quantum HSS will , in general, contain states that do not form full $N=2$
supermultiplets, such as $1/2$ BPS black holes etc...

We should also stress that the fundamental spectrum is  generated by the action of  analytic harmonic superfields involving an
infinite number of auxiliary fields. Once the auxiliary fields are eliminated the dynamical components of
the superfields become complicated nonlinear functions of the physical bosonic and fermionic fields.
Therefore the "fundamental spectrum" in HSS is in general not the simple Fock space of free bosons
and fermions.

The $N=2$ , $d=4$ MESGT's  under dimensional reduction lead to  $N=4$ , $d=3$ supersymmetric $
\sigma$ models with quaternionic K\"ahler manifolds $\mathcal{M}_3$ (C-map). After T-dualizing the three dimensional theory one can lift it back to four dimension, thereby obtaining an $N=2$ sigma model coupled to supergravity that is in the mirror image of the original $N=2$ MESGT. In previous sections we discussed how quantizing spherically symmetric
  stationary BPS black hole solutions of $4D$,  $N=2$ MESGTs defined by Jordan algebras naturally lead to  quaternionic discrete series representations of their three dimensional U-duality groups $QConf(J)$. The results summarized in this section suggest even a stronger result, namely,  quantized  solutions (spectra)  of a $4D$, $N=2$ MESGT , that allow dimensional reduction to three dimensions,  must fit into the minimal unitary representation of its three dimensional U-duality group and  those representations obtained by tensoring of minimal representations. 
  
\section{M/Superstring Theoretic Origins of $N=2$ MESGTs with Symmetric Scalar Manifolds}
\renewcommand{\theequation}{\arabic{section}.\arabic{equation}}
\setcounter{equation}{0}
Both the maximal supergravity and  the exceptional $N=2$ MESGT defined by the exceptional Jordan algebra $J_3^{\mathbb{O}}$ have exceptional groups of the $E$ series as their U-duality symmetry groups in five , four and three dimensions. The numbers of vector fields of these two theories are the same in five and four dimensions. However, the real forms of their U-duality groups  are different. In the Table \ref{excvsmaximal} we list the U-duality groups of the $N=2$ exceptional MESGT and those of maximal $N=8$ supergravity.
\begin{table}
\begin{tabular}{|c|c|c|c|} \hline
$d$ & $ J_3^\mathbb{O}$ MESGT  & $N=8$ Supergravity & $J_3^{\mathbb{H}}$ MESGT  \\ \hline
5 & $E_{6(-26)}$ & $E_{6(6)}$ & $SU^*(6)$ \\ \hline
4 & $E_{7(-25)}$ & $E_{7(7)}$ & $SO^*(12)$  \\ \hline
3 &$E_{8(-24)}$ &$E_{8(8)}$ & $E_{7(-5)}$ \\ \hline
\end{tabular}
\caption{ \label{excvsmaximal} Above we give the U-duality groups of the exceptional $N=2$ MESGT defined by the exceptional Jordan algebra and of the maximal $N=8$ supergravity in five, four and three spacetime dimensions. In the last column we list the U-duality group of the $N=2$ MESGT defined by $J_3^{\mathbb{H}}$ that is a common sector of these two theories. }
\end{table}
Just like the $N=8$ supergravity the fields of the exceptional $N=2$ MESGT could not be identified with  quarks and leptons so as to obtain a unified theory of all interactions without invoking a composite scenario \cite{Gunaydin:1983rk,Gunaydin:1984be}. However, it was observed that unlike the maximal $N=8$ supergravity theory one might be able to  couple matter multiplets to the exceptional supergravity theory which could be identified with quarks and leptons \cite{Gunaydin:1983rk,Gunaydin:1984be}.
These two theories have a common sector which is the $N=2$ MESGT defined by the quaternionic Jordan algebra $ J_3^{\mathbb{H}}$ \cite{Gunaydin:1983rk}. The existence of the exceptional MESGT begs the question whether there exists a larger theory that can be "truncated" to both the exceptional MESGT and the maximal supergravity theory \cite{Gunaydin:1984be}. After the discovery of Green-Schwarz anomaly cancellation and the development of Calabi-Yau technology this question evolved into the question whether one could obtain the  exceptional $4D$ and $5D$ MESGTs as low energy effective theories of type II superstring theory or M-theory on some exceptional Calabi-Yau manifold \cite{Gunaydin:1986fg}.  In fact more generally one would like to know if $N=2$ MESGTs with symmetric target spaces can be obtained from M/superstring theory on some Calabi-Yau manifold with or without hypermultiplet couplings. 
 The compactifications of M/Superstring theory over generic  Calabi-Yau manifolds do not  , in general, lead to symmetric or homogeneous scalar manifolds in the corresponding five or four dimensional  theories \cite{Candelas:1985en,Cadavid:1995bk}.  In the mathematics literature this was posed as the question whether  Hermitian symmetric spaces could arise as moduli spaces of deformations of Hodge structures of Calabi-Yau manifolds \cite{MR1258484}. In particular, the difficulty of obtaining the exceptional moduli space $\frac{E_{7(-25)}}{E_6\times U(1)}$ was highlighted by Gross~\cite{MR1258484}.

Using methods developed earlier in \cite{Vafa:1995gm}   Sen and Vafa  constructed dual pairs of  compactifications of type II superstring with $N=2,4$ and $N=6$ supersymmetries in $d=4$ by orbifolding $T^4 \times S^1 \times S^1$\cite{Sen:1995ff}.  Remarkably,   the low energy effective theory of one of the compactifications they study is  precisely the magical $N=2$ MESGT defined by the quaternionic Jordan algebra $J_3^{\mathbb{H}}$ without any hypermultiplets \cite{talkias,talkparis}. This follows from the fact that the resulting $N=2$ theory has the same bosonic field content as the $N=6$ supergravity which they also obtain via orbifolding.  As was shown in \cite{Gunaydin:1983rk} , the bosonic sector of $N=6$ supergravity and the $N=2$ MESGT defined by $J_3^{\mathbb{H}}$ are identical and their scalar  manifold in $d=4$ is $SO^*(12)/U(6)$ \footnote{ The authors of \cite{Sen:1995ff} appear to be unaware of this fact. It is easy to show that the resulting $N=2$ supergravity with 15 vector multiplets can not belong to the generic Jordan family with the scalar manifold $\frac{ SO(14,2)\times SU(1,1)}{SO(14)\times U(1)\times U(1)}$.}. This theory is self-dual with the dilaton belonging to a vector multiplet. In addition to the magical $N=2$ MESGT defined by $J_3^{\mathbb{H}}$, Sen and Vafa gave several other examples of compactifications with $N=2$ supersymmetry and symmetric target manifolds in $d=4$. One is the STU model coupled to 4 hypermultiplets with scalar manifold
\eq
\mathcal{M}_V \times \mathcal{M}_H = {[}\frac{SU(1,1)}{U(1)} {]}^3 \times \frac{SO(4,4)}{SO(4)\times SO(4)} 
\en
which is also self-dual. Another theory they construct leads to $N=2$ supergravity belonging to the generic Jordan family defined by the Jordan algebra $J=\mathbb{R} \oplus \Gamma_{(1,6)}$ with the target space
\[ \mathcal{M}_V = \frac{SO(6,2) \times SU(1,1)}{SO(6)\times U(1)\times U(1)} \]
which is not self-dual. 

A well-known theory with a symmetric target space that descends from type II string theory is the FSHV model \cite{Ferrara:1995yx}. It is obtained by compactification on a self-mirror Calabi-Yau manifold with  Hodge numbers $h^{(1,1)}=h^{(2,1)}=11$ and has the $4D$ scalar manifold
\eq
\mathcal{M}_V \times \mathcal{M}_H = \frac{SO(10,2) \times SU(1,1)}{SO(10)\times U(1)\times U(1) } \times \frac{SO(12,4)}{SO(12) \times SO(4)} 
\en
corresponding to the MESGT defined by the Jordan algebra $(\mathbb{R} \oplus \Gamma_{(1,9)})$ coupled to 12 hypermultiplets. 

It was known for  sometime  \cite{gunaydinsezgin} that there exists a six dimensional $(1,0)$ supergravity theory , which is free from gravitational anomalies , with 16 vector multiplets, 9 tensor multiplets and 28 hypermultiplets, that parametrize  the exceptional quaternionic symmetric  space $E_{8(-24)}/E_7 \times SU(2)$\footnote{ This theory remains anomaly free if one replaces the hypermultiplet sector with any $ 112$ ( real) dimensional quaternionic sigma model.}. This theory reduces to the exceptional supergravity theory defined by $J_3^{\mathbb{O}}$ coupled to the hypermultiplets in lower dimensions. It has the $4D$ scalar manifold
\[ \mathcal{M}_V \times \mathcal{M}_H = \frac{E_{7(-25)}}{E_6\times U(1)} \times \frac{E_{8(-24)}}{E_7\times SU(2)}  \] and in three dimensions  the moduli space becomes
 \eq
 \frac{E_{8(-24)}}{E_7\times SU(2)} \times \frac{E_{8(-24)}}{E_7\times SU(2)}
 \en
  The moduli space of the FSHV model is a subspace of this doubly exceptional moduli space. The existence of an anomaly free theory in $d=6$ that  reduces to this doubly exceptional theory whose moduli space includes that of the FSHV model suggests strongly that there must exist an exceptional Calabi-Yau manifold such that M/superstring theory compactified over it yields this doubly exceptional theory as was argued in \cite{talkias,talkparis}.   

Some important developments about the stringy origins of magical supergravity theories that took place after SAM 2007 is summarized in the next section.

\section{Recent Developments and Some Open Problems}
\setcounter{equation}{0}
Since they were delivered, a great deal  of progress has been made on the main  topics covered in these lecture, which I will try to summarize briefly in this section. 

We discussed at some length the proposals that four and three dimensional U-duality groups
act as spectrum generating conformal and quasiconformal groups of five
and four dimensional supergravity theories with symmetric scalar
manifolds respectively. Applying these proposals in succession  implies that three dimensional U-duality groups should act as spectrum generating 
symmetry groups of the five dimensional theories from which they descend. 
Several results that have been obtained recently lend further support to these proposals.   The authors  of \cite{Bouchareb:2007ax,Gal'tsov:2008nz} used 
solution generating techniques to relate the known black hole
solutions of simple five dimensional ungauged supergravity  and those  of the $5D$ STU model to each other and
generate new solutions by the action of corresponding $3D$   U-duality  groups $G_{2(2)}$ and $SO(4,4)$, respectively. 
For simple ungauged $5D$ supergravity similar results were also obtained in \cite{Compere:2009zh}  and for  gauged $5D$ supergravity  in \cite{Berkooz:2008rj}. 

Attractor flows  for 
non-BPS extremal black holes that are described by certain pseudo-Riemannian symmetric  sigma models coupled to $3D$ supergravity 
 was carried out in \cite{Gaiotto:2007ag}\footnote{Preliminary results of this work was reported by Li \cite{Li:2008ar}   in this school after these lectures were delivered.}.  Extremal BPS and non-BPS black holes of supergravity theories with symmetric target spaces were also studied in \cite{Bergshoeff:2008be,Bossard:2009at}. For a more complete up-to-date list of references  on extremal BPS and non-BPS attractors and their orbit structures I refer to \cite{ Bellucci:2009qv}.

As we discussed above the quantization of the attractor flows of stationary spherically symmetric $4D$ $N=2$ BPS  black holes 
leads to wave-functions that form the basis of quaternionic discrete series representations of the corresponding $3D$ U-duality groups 
$QConf(J)$. 
 Unitary representations of two quaternionic
groups of rank two, namely $SU(2,1)$ and $G_{2(2)}$, induced by
their geometric quasiconformal actions were studied in  \cite{Gunaydin:2007qq}.   The
starting point of the constructions of unitary representations given in \cite{Gunaydin:2007qq} are the spherical
vectors with respect to the  maximal compact subgroups  $SU(2)\times U(1)$ and $SU(2)\times SU(2)$ , respectively, under their quasiconformal
actions. In a recent paper with Pavlyk \cite{Gunaydin:2009dq}  we gave a  unified realization of  three dimensional U-duality groups of all  $N=2$ MESGTs defined by Euclidean Jordan algebras of degree three  as spectrum generating quasiconformal groups covariant with respect to their $5D$ U-duality groups.
The spherical vectors of quasiconformal realizations of all these groups twisted by a unitary character with respect to their maximal compact subgroups as well as  their quadratic Casimir operators and  their values were also given in \cite{Gunaydin:2009dq}. In a subsequent paper \cite{Gunaydin:2009zz}
we extended these results to  a unified realization of split exceptional groups $F_{4(4)},E_{6(6)}, E_{7(7)}, E_{8(8)}$ and of $ SO(n+3,m+3)$ as quasiconformal groups that is covariant with respect to their subgroups $SL(3,R), SL(3,R)\times SL(3,R), SL(6,R), E_{6(6)}$ and $SO(n,m)\times SO(1,1)$, respectively, and determined their spherical vectors.  We also gave their quadratic Casimir operators and determined their values in terms of $\nu$ and the dimension $n_V$ of the underlying Jordan algebras. For $\nu= -(n_V+2)+i\rho$ the quasiconformal action induces unitary representations on the space of square integrable functions in $(2n_V+3)$ variables, that belong to the principle series. For special discrete values of $\nu$ the quasiconformal action leads to unitary representations belonging to the discrete series and their continuations.  For the quaternionic real forms of $3D$ U-duality groups these discrete series representations belong to the quaternionic discrete series. 

As I discussed in section 11, the minimal unitary representations of non-compact groups are the analogs of the singleton representations of the symplectic groups $Sp(2n, \mathbb{R})$. Now $Sp(4,\mathbb{R})$  is the covering group  $4D$ anti-de Sitter or $3D$ conformal group $SO(3,2)$.  For higher AdS or conformal groups  the analogous  representations were referred to as doubletons \cite{Gunaydin:1984wc,Gunaydin:1984fk}.  Singleton or doubleton representations of AdS or conformal groups $SO(d,2)$ are singular positive energy ( lowest weight) unitary representations , which are in the continuation of holomorphic discrete representations. Even though they do not belong to the holomorphioc discrete series , by tensoring singletons or doubletons one can obtain  the entire holomorphic discrete series representations of AdS or conformal groups.  This fact lies at the heart of AdS/CFT dualities \cite{Maldacena:1997re}. In fact, the entire K-K spectrum  of IIB supergravity over $AdS_5\times S^5$ was obtained by tensoring the CPT invariant doubleton supermultiplet repeatedly with itself in
\cite{Gunaydin:1984fk}. Again it was pointed out in \cite{Gunaydin:1984fk} that  the doubleton supermultiplet decouples from the K-K spectrum as gauge modes and its field theory is the conformally invariant  $N=4$ super Yang-Mills theory  that lives on the boundary of $AdS_5$ which is the $4D$ Minkowski space. Similar results were obtained  for  11-dimensional supergravity on $AdS_4 \times S^7$ \cite{Gunaydin:1985tc} and on $AdS_7 \times S^4$ \cite{Gunaydin:1984wc}.  

Similarly, the minimal unitary representations of noncompact groups of  quaternionic type  do not belong to the quaternionic discrete series , but are in their singular continuations \cite{MR1421947}.  I argued in section 11  that the "fundamental spectrum" of $N=2$ sigma models coupled to supergravity with symmetric target manifolds $ G/K\times SU(2)$  must belong to the minimal unitary representation of $G$ obtained by quantization of the geometric quasiconformal realization of $G$. On the other hand quantization of the geodesic 
motion describing the dynamics of spherically symmetric BPS black holes lead to wave functions belonging to the quaternionic discrete series representations of $G$ \cite{Gunaydin:2007qq}. This implies that by tensoring minimal unitary representations one should be able to obtain the quaternionic discrete series representations. Decomposition of tensor products of minimal unitary representations into its irreducible  pieces is much harder for general noncompact groups since their minimal representations are  , in general, not of the lowest weight type and their realizations are non-linear ( interacting)! This problem is currently under investigation\cite{mg2009}.

Several novel results were obtained regarding the stringy origins of magical supergravity theories since SAM 2007.  In \cite{Dolivet:2007sz} the hyper-free $N=2$ string models based on asymmetric orbifolds  with 
$N=(4,1)$ world-sheet superconformal symmetry using $2D$  fermionic construction were given. Among these models two of them correspond to two of the magical supergravity theories, namely the $J_3^{\mathbb{C}}$  
MESGT with the  moduli space
\[ \mathcal{M}_4= \frac{SU(3,3)}{SU(3) \times SU(3) \times U(1)} \]
and the $J_3^{\mathbb{H}}$ theory with the moduli space
\[ \mathcal{M}_4= \frac{SO^*(12)}{U(6)} \]

In \cite{Bianchi:2007va},
Bianchi and Ferrara  reconsidered the string derivation of  FSHV model over the Enriques Calabi-Yau and argued that the exceptional supergravity theory defined by octonionic $J_3^\mathbb{O}$ admits a string interpretation closely related to the Enriques model. They show that the  uplift of the exceptional MESGT to $d=6$ has 16 Abelian vectors which is related to the rank of Type I and heterotic strings. 

In mathematics literature, Todorov gave a construction of Calabi-Yau n-folds whose moduli spaces are locally symmetric spaces \cite{todorov-2008}. This family of CY manifolds of complex dimension $n$  are double covers of the projective $n$ dimensional spaces ramified over $2n+2$ hyperplanes. They have the Hodge numbers 
\[ h^{(n-p,p)} = \left(\begin{array}{c} n \\ p\end{array}\right)^2 \]
and have the second Betti numbers
\[ {b}_2= \left(\begin{array}{c}(2n+2) \\2\end{array}\right) +1 \]
 Type IIB  superstring theory compactified over such a CY threefold constructed by Todorov leads to the magical $N=2$ MESGT defined by the complex Jordan algebra $J_3^{\mathbb{C}}$ with vector moduli space
\[ \mathcal{M}_4 = \frac{SU(3,3)}{SU(3)\times SU(3)\times U(1)} \]
coupled to $ ( h^{(1,1)}+1 ) =30 $ hypermultiplets. Todorov promises a sequel to his paper in which he would show that there are no instanton corrections to the moduli space of this  Calabi-Yau manifold.

The fact that some $N=2$  MESGTs  with symmetric target spaces , in particular the magical supergravity theories, can be obtained from M/Superstring theory is of utmost importance. It implies that the corresponding supergravity theories have quantum completions and  discrete arithmetic subgroups of their U-duality groups will survive at the non-perturbative regime of M/Superstring theory \cite{Hull:1994ys}. The relevant unitary representations of the spectrum generating symmetry groups will then be some " automorphic representations".  If and  how some of the results obtained  for continuous U-duality groups extend to their discrete   subgroups is a wide  open problem. 

Another open problem is the quantization of $4D$, $N=2$ sigma models that couple to supergravity in harmonic superspace and their lower dimensional descendents.

{ \bf Acknowledgements: }
I would like to thank Stefano Bellucci for his kind invitation to deliver these lectures at  SAM 2007 and the participants for numerous stimulating discussions. I wish to thank Stefano Bellucci, Sergio Ferrara, Kilian Koepsell, Alessio Marrani, Andy Neitzke, Hermann Nicolai, Oleksandr Pavlyk ,  Boris Pioline and Andrew Waldron for enjoyable collaborations and stimulating discussions on various topics covered
in these lectures. 
Thanks are also due to the organizers of the "Fundamental Aspects of Superstring Theory 2009" Workshop at KITP, UC Santa Barbara and of the 
  New Perspectives in String Theory 2009 Workshop at GGI, Florence where part of these lectures were written up.
  This work was supported in part by the National
Science Foundation under grant number PHY-0555605. Any opinions,
findings and conclusions or recommendations expressed in this
material are those of the authors and do not necessarily reflect the
views of the National Science Foundation.

\providecommand{\href}[2]{#2}\begingroup\raggedright\endgroup

\end{document}